\documentclass[12pt]{article}
\usepackage{graphicx}
\usepackage{amsmath,amssymb,amsfonts,amsthm}
\usepackage{setspace}
\usepackage{subcaption}
\usepackage{booktabs}
\usepackage{multirow}
\usepackage{hyperref}
\hypersetup{
    colorlinks,%
    citecolor=black,%
    filecolor=black,%
    linkcolor=black,%
    urlcolor=blue
}
\numberwithin{equation}{section}

\usepackage[top=1.25 in, bottom=1.25 in, left=1.25 in , right=1.25 in]{geometry}
\usepackage{titlesec}
\titleformat{\section}{\large\bfseries}{\thesection}{1em}{}
\titleformat{\subsection}{\normalsize\bfseries}{\thesubsection}{1em}{}

\usepackage{enumitem}
\setdescription{leftmargin=\parindent,labelindent=\parindent}

\newtheorem{proposition}{Proposition}

\newenvironment{definition}[1][Definition]{\begin{trivlist}
\item[\hskip \labelsep {\bfseries #1}]}{\end{trivlist}}

\newtheorem{remark}{Remark}

\begin{document}

\title{\Large{A Consistent Variance Estimator for 2SLS\\ When Instruments Identify Different LATEs}}
\author{Seojeong Lee\footnote{School of Economics, UNSW Business School, University of New South Wales, Sydney NSW 2052 Australia, email: \href{mailto:jay.lee@unsw.edu.au}{jay.lee@unsw.edu.au}, web: \href{https://sites.google.com/site/misspecifiedjay/}{https://sites.google.com/site/misspecifiedjay/}}
\footnote{Conversations with Tue Gorgens motivated this paper. I also thank Bruce Hansen, Jack Porter, Han Hong, Toru Kitagawa, Sukjin Han, Ju Hyun Kim, Gigi Foster, Valentyn Panchenko, Denzil Fiebig, Youngki Shin, Taehoon Kim, and Peter Siminski as well as session participants at SETA 2014, KAEA-KEA, the Econometric Society World Congress 2015 conferences for their helpful discussions and suggestions.}}

\date{Accepted for publication at the Journal of Business \& Economic Statistics}
\maketitle

\vspace{-2em}

\begin{abstract}
Under treatment effect heterogeneity, an instrument identifies the instrument-specific local average treatment effect (LATE). With multiple instruments, two-stage least squares (2SLS) estimand is a weighted average of different LATEs. What is often overlooked in the literature is that the postulated moment condition evaluated at the 2SLS estimand does not hold unless those LATEs are the same. If so, the conventional heteroskedasticity-robust variance estimator would be inconsistent, and 2SLS standard errors based on such estimators would be incorrect. I derive the correct asymptotic distribution, and propose a consistent asymptotic variance estimator by using the result of Hall and Inoue (2003, \textit{Journal of Econometrics}) on misspecified moment condition models. This can be used to correctly calculate the standard errors regardless of whether there is more than one LATE or not.\\

\end{abstract}

\section{Introduction}

Since the series of seminal papers by Imbens and Angrist (1994), Angrist and Imbens (1995), and Angrist, Imbens, and Rubin (1996), the local average treatment effect (LATE) has played an important role in providing useful guidance to many policy questions. The key underlying assumption is treatment effect heterogeneity, i.e. each individual has a different causal effect of treatment on outcome. Assume a binary treatment, $D_{i}$, and an outcome variable $Y_{i}$. Let $Y_{1i}$ and $Y_{0i}$ denote the potential outcomes of individual $i$ with and without the treatment, respectively. The heterogeneous individual treatment effect is $Y_{1i}-Y_{0i}$, but this cannot be identified because $Y_{1i}$ and $Y_{0i}$ are never observed at the same time. Therefore, researchers typically focus on the average treatment effect (ATE), $E[Y_{1i}-Y_{0i}]$. However, unless the treatment is randomly assigned, a naive estimate of ATE is biased because of selection into treatment. 

Instrumental variables are used to overcome this endogeneity problem. If an instrument $Z_{i}$ which is randomly assigned, independent of $Y_{1i}$ and $Y_{0i}$, and correlated with the treatment $D_{i}$ is available, then ATE of those whose treatment status can be changed by the instrument, thus the \textit{local} ATE, is identified. Assume $Z_{i}$ is binary and define $D_{1i}$ and $D_{0i}$ be $i$'s treatment status when $Z_{i}=1$ and $Z_{i}=0$, respectively. The LATE theorem of Imbens and Angrist (1994) shows that 
\begin{equation}
\label{LATE}
\frac{Cov(Y_{i},Z_{i})}{Cov(D_{i},Z_{i})}=\frac{E[Y_{i}|Z_{i}=1]-E[Y_{i}|Z_{i}=0]}{E[D_{i}|Z_{i}=1]-E[D_{i}|Z_{i}=0]}= E[Y_{1i}-Y_{0i}|D_{1i}>D_{0i}].
\end{equation}
That is, the instrumental vaiables (IV) estimand (or the Wald estimand) is equal to the ATE of those with $D_{1i}>D_{0i}$, who are called compliers. Those who take the treatment regardless of the instrument value, $D_{1i}=D_{0i}=1$, are always-takers, and those who do not take the treatment anyway, $D_{1i}=D_{0i}=0$, are never-takers. We cannot identify ATE of always-takers and never-takers in general. By the monotonicity assumption of Imbens and Angrist, we exclude defiers who behave in the opposite way with compliers, $D_{1i}<D_{0i}$. Since the compliers are specific to the instrument $Z_{i}$, LATE is instrument-specific.

The above setting can be generalized to situations where the number of (excluded) instruments is greater than the number of endogenous variables. The two-stage least squares (2SLS) estimator is commonly used to estimate the causal effect in such cases. Without loss of generality, consider mutually exclusive binary instruments, $Z_{i}^{j}$ for $j=1,...,q$. Let $D_{zi}^{j}$ be $i$'s potential treatment status when $Z_{i}^{j}=z$ where $z=0,1$. Each instrument identifies a version of LATE because compliers may differ for each $Z_{i}^{j}$. The 2SLS estimand is a weighted average of treatment effects for the instrument-specific compliers:
\begin{equation}
\rho_{a} = \sum_{j=1}^{q}\xi_{j}\cdot E[Y_{1i}-Y_{0i}|D_{1i}^{j}>D_{0i}^{j}],
\end{equation}
where $0\leq\xi_{j}\leq1$ and $\sum_{j}\xi_{j}=1$. This is first shown by Imbens and Angrist (1994) for a discrete instrument which can be written as mutually exclusive binary instruments. Angrist and Imbens (1995) and Heckman and Vytlacil (2005) extend this result to multi-valued treatment with covariates, and a continuous instrument with covariates, respectively. These works provided theoretical foundations to interpret 2SLS point estimates as a weighted average of LATEs, and empirical researchers have done so, either explicitly or implicitly.

If the 2SLS estimand is a weighted average of more than one LATE, then the over-identifying restrictions test (the J test, hereinafter) which is typically conducted along with 2SLS would reject the null hypothesis that the moment condition is correctly specified.\footnote{The J test can also reject the null due to invalid instruments. Kitagawa (2015) proposes a specification test for instrument validity under treatment effect heterogeneity.} What is less well known and often overlooked in the literature is that the conventional standard errors are no longer correct under misspecification of the moment condition. This fact has been neglected and the standard errors have been routinely calculated assuming the LATEs are identical. I derive the asymptotic distribution of 2SLS when the estimand is a weighted average of LATEs and propose a consistent estimator of the asymptotic variance robust to multiple LATEs. The correct standard error based on the proposed variance estimator (the multiple-LATEs-robust standard error, hereinafter) can be substantially different from the conventional heteroskedasticity-robust one even for a large sample size, or even for p-values above any usual significance level. 

Two recent papers cover similar topics. Koles\'{a}r (2013) shows that under treatment effect heterogeneity the 2SLS estimand is a convex combination of LATEs while the limited information maximum likelihood (LIML) estimand may not. Angrist and Fernandez-Val (2013) propose an estimand for new subpopulations by reweighting covariate-specific LATEs. However, neither of the two papers considers correct variance estimation of 2SLS.

In the next Section, I show that the postulated moment condition of 2SLS is misspecified when there is more than one LATE. The asymptotic distribution of 2SLS estimators in such a case is derived, and a consistent variance estimator is proposed. In Section \ref{weighted}, I discuss practical implications of using 2SLS with multiple instruments. In particular, Angrist and Krueger (1991), Angrist and Evans (1998), and Thornton (2008) are replicated and the originally reported conventional standard errors are compared to the multiple-LATE-robust standard errors. Section \ref{simulation} presents simulation results that show (i) the multiple-LATEs-robust standard error is closer to the standard deviation of the 2SLS estimator than the conventional standard error, (ii) the bias in the conventional standard error tends to be large when the first stage F statistic or the p-value of the J statistic are small, and (iii) the conclusion of $t$ tests may change if the multiple-LATEs-robust standard error is used regardless of the magnitude of the bias in the conventional standard error. Section \ref{conclusion} concludes. The proofs of propositions are collected in the appendix.

\section{Moment condition for 2SLS}
\label{mc}
I first define moment condition models for IV and 2SLS estimators to derive the asymptotic distribution. I maintain the assumption that the treatment variable and instruments are binary for simplicity of exposition, but this will be relaxed later in this section. The observed outcome can be written as
\begin{equation}
Y_{i} =Y_{1i}D_{i} + Y_{0i}(1-D_{i})=E[Y_{0i}] + D_{i}\rho_{i} + \eta_{i},
\label{eq1}
\end{equation}
where $\rho_{i}=Y_{1i}-Y_{0i}$ and $\eta_{i}=Y_{0i}-E[Y_{0i}]$. Since the individual treatment effect $\rho_{i}$ cannot be identified, a version of ATE becomes the parameter of interest. Let $\rho$ be the parameter, and $\alpha$ be a nuisance parameter for the intercept. Rewriting \eqref{eq1} in a familiar regression form using $\alpha$ and $\rho$, we get
\begin{eqnarray}
\label{model2sls}
Y_{i} &=& \alpha + D_{i}\rho + e_{i},\\
\nonumber e_{i} &\equiv&e_{i}(\alpha,\rho)= E[Y_{0i}] - \alpha + D_{i}(\rho_{i}-\rho) + \eta_{i}.
\end{eqnarray}
It is straightforward to see that regressing $Y_{i}$ on a constant and $D_{i}$ yields
\begin{equation}
\alpha_{ols} = E[Y_{i}]-E[D_{i}]\cdot\rho_{ols},\hspace{1em}\rho_{ols} = \frac{Cov(Y_{i},D_{i})}{Var(D_{i})},
\end{equation}
which are the solutions to the moment condition 
\begin{eqnarray}
\label{mc_ols}
0&=& E[e_{i}^{ols}] = E[Y_{i}-\alpha_{ols}-D_{i}\rho_{ols}],\\
\nonumber 0 &=& E[D_{i}e_{i}^{ols}]=E[D_{i}(Y_{i}-\alpha_{ols}-D_{i}\rho_{ols})].
\end{eqnarray}
Although the moment condition \eqref{mc_ols} is satisfied at $(\alpha_{ols},\rho_{ols})$, $\rho_{ols}$ does not identify an interesting population parameter (the ATE on the treated, in this case) because $D_{i}$ is endogenous, i.e. $Y_{0i}\not\perp D_{i}$, and
\begin{eqnarray}
\rho_{ols} =\frac{Cov(Y_{i},D_{i})}{Var(D_{i})} &=& E[Y_{i}|D_{i}=1]-E[Y_{i}|D_{i}=0]\\
\nonumber &=&E[\rho_{i}|D_{i}=1] + E[Y_{0i}|D_{i}=1]-E[Y_{0i}|D_{i}=0].
\end{eqnarray}
Now suppose that there is a binary instrument $Z_{i}^{1}$ which satisfies Conditions 1-3 of Imbens and Angrist (1994). The IV estimator is based on the moment condition
\begin{eqnarray}
\label{mc_iv}
0&=& E[e_{i}^{1}] = E[Y_{i}-\alpha_{IV}^{1}-D_{i}\rho_{IV}^{1}],\\
\nonumber 0 &=& E[Z_{i}^{1}e_{i}^{1}]=E[Z_{i}^{1}(Y_{i}-\alpha_{IV}^{1}-D_{i}\rho_{IV}^{1})],
\end{eqnarray}
where the unique solution $(\alpha_{IV}^{1},\rho_{IV}^{1})$ is given by
\begin{eqnarray}
\label{alphaIV}
\alpha_{IV}^{1} &=& E[Y_{i}]-E[D_{i}]\cdot \rho_{IV}^{1},\\
\rho_{IV}^{1} &=& \frac{Cov(Y_{i},Z_{i}^{1})}{Cov(D_{i},Z_{i}^{1})} = E[\rho_{i}|D_{1i}^{1}>D_{0i}^{1}].
\label{rhoIV}
\end{eqnarray}
The second equality of \eqref{rhoIV} holds by Theorem 2 of Imbens and Angrist (1994). For OLS or IV moment conditions, the model is just-identified because the number of unknown parameters is equal to the number of equations in the moment condition. This implies that there always exists a solution that makes the moment condition equal to zero.\footnote{This means that $\rho_{ols}$ should be interpreted as the projection coefficient. For IV, \eqref{mc_iv} holds even if $Z_{i}^{1}$ is not independent of $Y_{0i}$, and thus $E[Z_{i}^{1}\eta_{i}]\neq0$. In this case, the second equality of \eqref{rhoIV} does not hold. In other words, $0=E[Z_{i}^{1}e_{i}^{1}]$ is not the instrument validity condition under treatment effect heterogeneity.} The asymptotic distributions of the OLS and IV estimators are derived by expanding the first-order conditions (FOC) around their estimands and using the fact that the respective moment condition holds. Thus, the conventional standard errors are correct under treatment effect heterogeneity.

This positive conclusion does not hold if there are more instruments than the endogenous parameters. In this case, the moment condition is over-identified and the assumption that there exists a unique solution to the moment condition may be violated. Suppose that there are two valid instruments, $Z_{i}^{1}$ and $Z_{i}^{2}$. If we use each instrument one at a time, we would get $(\alpha_{IV}^{1},\rho_{IV}^{1})$ and $(\alpha_{IV}^{2},\rho_{IV}^{2})$, where each corresponds to a different LATE. Assume $\rho_{IV}^{1}\neq\rho_{IV}^{2}$. I emphasize that they are the unique solutions to the corresponding moment conditions. The 2SLS estimator that uses both instruments at the same time is calculated by estimating $D_{i} = \delta + Z_{i}^{1}\pi_{1} + Z_{i}^{2}\pi_{2} + u_{i}$ by OLS in the first stage to produce the fitted value $\hat{D}_{i}$, and then by estimating $Y_{i} = \alpha +  \hat{D}_{i}\rho + e_{i}$ by OLS in the second stage. This estimator is equivalent to the linear GMM using the sample mean of outer-products of the instrument vector as a weight matrix, based on the moment condition
\begin{eqnarray}
\label{mc_2sls}
0&=& E[e_{i}^{*}] = E[Y_{i}-\alpha^{*}-D_{i}\rho^{*}],\\
\nonumber 0 &=& E[Z_{i}^{1}e_{i}^{*}]=E[Z_{i}^{1}(Y_{i}-\alpha^{*}-D_{i}\rho^{*})],\\
\nonumber 0 &=& E[Z_{i}^{2}e_{i}^{*}]=E[Z_{i}^{2}(Y_{i}-\alpha^{*}-D_{i}\rho^{*})],
\end{eqnarray}
for a unique parameter vector $(\alpha^{*},\rho^{*})$. To see if there exists such a solution, solve the first equation for $\alpha^{*}$, substitute it into the second and third equations, and solve them for $\rho^{*}$ to have
\begin{eqnarray}
\label{alpha*}
\alpha^{*} &=& E[Y_{i}]-E[D_{i}]\cdot \rho^{*},\\
\rho^{*} &=& \frac{Cov(Y_{i},Z_{i}^{1})}{Cov(D_{i},Z_{i}^{1})} = \frac{Cov(Y_{i},Z_{i}^{2})}{Cov(D_{i},Z_{i}^{2})}.
\label{rho*}
\end{eqnarray}
By the LATE theorem, \eqref{rho*} implies that $\rho^{*}=\rho_{IV}^{1}=\rho_{IV}^{2}$, but this contradicts the assumption that $\rho_{IV}^{1}\neq\rho_{IV}^{2}$. In other words, if $(\alpha_{IV}^{1},\rho_{IV}^{1})$ were the unique solution to the first two equations of \eqref{mc_2sls}, then it should satisfy the last equation, but this implies that the two LATEs are the same. Thus, there does not exist a unique parameter that satisfies \eqref{mc_2sls} and the moment condition is misspecified. It is stressed that misspecification of the moment condition does not imply invalidity of the instruments under treatment effect heterogeneity. 
\begin{definition}[Definition 1 (Hall and Inoue, 2003)]
	A model is said to be \textit{misspecified} if there is no value of $\theta$ which satisfies the assumed over-identified moment condition.
\end{definition}
This definition is not directly related to functional form or data-generating process misspecification. For instance, the FOC of the quasi-maximum likelihood forms a just-identified moment condition with a unique solution. Therefore, it is a correctly specified moment condition, although the distribution is misspecified. When the moment condition is misspecified, the GMM estimator is consistent for a value that minimizes the population criterion function. This value is called the pseudo-true value, and it is the 2SLS estimand in this context. Imbens and Angrist (1994) show that the 2SLS estimand is
\begin{eqnarray}
\label{alpha2sls}
\alpha_{0} &=& E[Y_{i}] - E[D_{i}]\cdot \rho_{0},\\
\rho_{0} &=& \xi\cdot\rho_{IV}^{1} + (1-\xi)\cdot\rho_{IV}^{2},\hspace{1em}0\leq\xi\leq1.
\label{rho2sls}
\end{eqnarray}
Evaluated at $(\alpha_{0},\rho_{0})$, the moment function \eqref{mc_2sls} does not hold. This conclusion continues to hold with more than two instruments and covariates.

Misspecified moment conditions under treatment effect heterogeneity have important implications. First, the J test will reject the null hypothesis \eqref{mc_2sls} asymptotically. It is not surprising that researchers often face a significant J test statistic when multiple instruments are used. If we can rule out the possibility of invalid instruments either by a statistical test such as Kitagawa (2015) or by an economic reasoning, the rejection is due to treatment effect heterogeneity. Thus, the J test has little relevance once heterogeneity is already assumed. Second, the asymptotic variance of 2SLS will be different from the standard one, and the conventional heteroskedasticity-robust variance estimator would be inconsistent. It is surprising that this has been overlooked in the literature. In the following propositions, I derive the asymptotic distribution of 2SLS and propose a consistent variance estimator. This variance estimator should always be used to calculate the standard error of 2SLS, regardless of the J test results.



To formally derive the asymptotic distribution, I consider the model \eqref{model2sls} with covariates where the treatment variable and instruments can take multiple values or even be continuous as in Angrist and Imbens (1995) and Heckman and Vytlacil (2005). Although I assume a single endogenous variable, extensions to multiple endogenous variables are straightforward. Assume that there are valid instruments $Z_{i}^{1},Z_{i}^{2},...,Z_{i}^{q}$. Let $(Y_{i},\mathbf{X}_{i},\mathbf{Z}_{i})_{i=1}^{n}$ be an iid sample, where $\mathbf{X}_{i}=(\mathbf{W}_{i}',D_{i})'$, $\mathbf{Z}_{i}=(\mathbf{W}_{i}',Z_{i}^{1},\cdots,Z_{i}^{q})'$, and $\mathbf{W}_{i}$ be an $l\times1$ vector of covariates including a constant. The first and second stages are
\begin{eqnarray}
Y_{i} &=& \mathbf{W}_{i}'\boldsymbol{\gamma}+D_{i}\rho+e_{i}\equiv \mathbf{X}_{i}'\boldsymbol{\beta}+e_{i},\\
D_{i} &=& \mathbf{W}_{i}'\boldsymbol{\delta} + \mathbf{Z}_{i}'\boldsymbol{\pi} + u_{i},
\end{eqnarray}
where $\boldsymbol{\beta} = (\boldsymbol{\gamma}',\rho)'$, and $\boldsymbol{\gamma}$, $\boldsymbol{\delta}$, and $\boldsymbol{\pi}$ are conformable parameters. The 2SLS estimator is
\begin{equation}
\boldsymbol{\hat{\beta}} = (\mathbf{X}'\mathbf{Z}(\mathbf{Z}'\mathbf{Z})^{-1}\mathbf{Z}'\mathbf{X})^{-1}\mathbf{X}'\mathbf{Z}(\mathbf{Z}'\mathbf{Z})^{-1}\mathbf{Z}'\mathbf{Y},
\label{2sls}
\end{equation}
where $\mathbf{X}\equiv (\mathbf{X}_{1},\cdots,\mathbf{X}_{n})'$ is an $n\times (l+1)$ matrix, $\mathbf{Z}\equiv (\mathbf{Z}_{1},\cdots,\mathbf{Z}_{n})'$ is an $n\times (l+q)$ matrix, and $\mathbf{Y} \equiv (Y_{1},\cdots,Y_{n})'$ is an $n\times 1$ vector. The following proposition establishes the asymptotic distribution of 2SLS estimators when there is more than one LATE in the general setting.

\begin{proposition}
Let $\boldsymbol{\beta_{0}}=(\boldsymbol{\gamma_{0}}',\rho_{0})'$ be the 2SLS estimand where $\boldsymbol{\gamma_{0}}$ satisfies $E[Y_{i}]=E[\mathbf{W}_{i}]'\boldsymbol{\gamma_{0}}+E[D_{i}]\rho_{0}$ and $\rho_{0}$ is a linear combination of different LATEs. Let $e_{i}\equiv Y_{i}-\mathbf{X}_{i}'\boldsymbol{\beta_{0}}$. The asymptotic distribution of 2SLS is
\[\sqrt{n}(\boldsymbol{\hat{\beta}}-\boldsymbol{\beta_{0}})\overset{d}{\rightarrow}N(\mathbf{0},\mathbf{H}^{-1}\boldsymbol{\Omega} \mathbf{H}^{-1}),\]
where $\mathbf{H}=E[\mathbf{X}_{i}\mathbf{Z}_{i}']\left(E[\mathbf{Z}_{i}\mathbf{Z}_{i}']\right)^{-1}E[\mathbf{Z}_{i}\mathbf{X}_{i}']$, $\boldsymbol{\Omega}=E[\boldsymbol{\psi}_{i}\boldsymbol{\psi}_{i}']$, and
\begin{eqnarray}
\nonumber \boldsymbol{\psi_{i}} &=& E[\mathbf{X}_{i}\mathbf{Z}_{i}'](E[\mathbf{Z}_{i}\mathbf{Z}_{i}'])^{-1}\left(\mathbf{Z}_{i}e_{i}-E[\mathbf{Z}_{i}e_{i}]\right) + (\mathbf{X}_{i}\mathbf{Z}_{i}'-E[\mathbf{X}_{i}\mathbf{Z}_{i}'])(E[\mathbf{Z}_{i}\mathbf{Z}_{i}'])^{-1}E[\mathbf{Z}_{i}e_{i}]\\
\nonumber && + E[\mathbf{X}_{i}\mathbf{Z}_{i}'](E[\mathbf{Z}_{i}\mathbf{Z}_{i}'])^{-1}\left(E[\mathbf{Z}_{i}\mathbf{Z}_{i}']-\mathbf{Z}_{i}\mathbf{Z}_{i}'\right)\left(E[\mathbf{Z}_{i}\mathbf{Z}_{i}']\right)^{-1}E[\mathbf{Z}_{i}e_{i}].
\end{eqnarray}
\end{proposition}

The next proposition proposes a consistent estimator of the asymptotic variance matrix of 2SLS robust to multiple-LATEs.
\begin{proposition}
A multiple-LATEs-robust asymptotic variance estimator given by
\begin{equation}
\label{SigmaMR}
\boldsymbol{\hat{\Sigma}_{MR}} = \hat{\mathbf{H}}^{-1}\left(\frac{1}{n}\sum_{i}\boldsymbol{\hat{\psi}}_{i}\boldsymbol{\hat{\psi}}_{i}'\right)\hat{\mathbf{H}}^{-1}
\end{equation}
where $\hat{\mathbf{H}}=\frac{1}{n}\mathbf{X}'\mathbf{Z}\left(\frac{1}{n}\mathbf{Z}'\mathbf{Z}\right)^{-1}\frac{1}{n}\mathbf{Z}'\mathbf{X}$,
\begin{eqnarray}
\boldsymbol{\hat{\psi}}_{i} &=& \frac{1}{n}\mathbf{X}'\mathbf{Z}\left(\frac{1}{n}\mathbf{Z}'\mathbf{Z}\right)^{-1}\left(\mathbf{Z}_{i}\hat{e}_{i} -\frac{1}{n}\mathbf{Z}'\boldsymbol{\hat{e}}\right) \\
\nonumber && + \left(\mathbf{X}_{i}\mathbf{Z}_{i}'-\frac{1}{n}\mathbf{X}'\mathbf{Z}\right)\left(\frac{1}{n}\mathbf{Z}'\mathbf{Z}\right)^{-1}\frac{1}{n}\mathbf{Z}'\boldsymbol{\hat{e}} \\
\nonumber && + \frac{1}{n}\mathbf{X}'\mathbf{Z}\left(\frac{1}{n}\mathbf{Z}'\mathbf{Z}\right)^{-1}\left(\frac{1}{n}\mathbf{Z}'\mathbf{Z}-\mathbf{Z}_{i}\mathbf{Z}_{i}'\right)\left(\frac{1}{n}\mathbf{Z}'\mathbf{Z}\right)^{-1}\frac{1}{n}\mathbf{Z}'\boldsymbol{\hat{e}},
\end{eqnarray}
$\hat{e}_{i} = Y_{i}-\mathbf{X}_{i}'\boldsymbol{\hat{\beta}}$, and $\boldsymbol{\hat{e}}=(\hat{e}_{1},\hat{e}_{2},...,\hat{e}_{n})'$, is consistent for $\mathbf{H}^{-1}\boldsymbol{\Omega} \mathbf{H}^{-1}$.
\end{proposition}

The formula of $\boldsymbol{\hat{\Sigma}_{MR}}$ is different from that of the conventional heteroskedasticity-robust variance estimator:
\begin{equation}
\boldsymbol{\hat{\Sigma}_{C}} = \hat{\mathbf{H}}^{-1}\left(\frac{1}{n}\mathbf{X}'\mathbf{Z}\right)\left(\frac{1}{n}\mathbf{Z}'\mathbf{Z}\right)^{-1}\left(\frac{1}{n}\sum_{i}\mathbf{Z}_{i}\mathbf{Z}_{i}'\hat{e}_{i}^{2}\right)\left(\frac{1}{n}\mathbf{Z}'\mathbf{Z}\right)^{-1}\left(\frac{1}{n}\mathbf{Z}'\mathbf{X}\right)\hat{\mathbf{H}}^{-1}.
\end{equation}
Under constant treatment effects, both $\boldsymbol{\hat{\Sigma}_{MR}}$ and $\boldsymbol{\hat{\Sigma}_{C}}$ have the same probability limit, but they are generally different in finite samples. $\boldsymbol{\hat{\Sigma}_{MR}}$ is consistent for the true asymptotic variance matrix even when the postulated moment condition is misspecified, and thus can be used regardless of whether there is one or more than one LATE. In contrast, $\boldsymbol{\hat{\Sigma}_{C}}$ is consistent only if the underlying LATEs are identical. This is also true for the standard errors based on $\boldsymbol{\hat{\Sigma}_{MR}}$ and $\boldsymbol{\hat{\Sigma}_{C}}$.

When there is a single endogenous variable without covariates, Proposition 1 coincides with the result in the proof of Theorem 3 of Imbens and Angrist (1994) when the first stage is known but needs to be estimated.\footnote{There are typos in the proof of Theorem 3 of Imbens and Angrist (1994). Their matrix $\Delta$ shoud read $\Delta = \left(\begin{array}{ccc} E[\psi(Z,D,\theta)\cdot\psi(Z,D,\theta)'] & E[\varepsilon\cdot\psi(Z,D,\theta)] & E[g(Z)\cdot\varepsilon\cdot\psi(Z,D,\theta)] \\ E[\varepsilon\cdot\psi(Z,D,\theta)]' & E[\varepsilon^{2}] & E[g(Z)\cdot\varepsilon^{2}] \\ E[g(Z)\cdot\varepsilon\cdot\psi(Z,D,\theta)]' & E[g(Z)\cdot\varepsilon^{2}] & E[g^{2}(Z)\cdot\varepsilon^{2}] \end{array}\right)$.} Their derivation is based on the stacked moment condition that consists of FOCs of the first and second stages, which is a special case of two-step estimators of Newey and McFadden (1994). They use the condition that the population fitted value of the endogenous variable is uncorrelated with $e_{i}$, where $e_{i}$ is defined in Proposition 1. In other words, the estimated first stage is used as an instrument. For example, the condition is $E[(\delta+\pi_{1}Z_{i}^{1}+\pi_{2}Z_{i}^{2})e_{i}]=0$ for a two instruments case, which does not necessarily imply $E[Z_{i}^{1}e_{i}]=E[Z_{i}^{2}e_{i}]=0$. This makes their asymptotic variance and its estimator robust to violations of the underlying moment condition of 2SLS, $E[Z_{i}^{1}e_{i}]=E[Z_{i}^{2}e_{i}]=0$. Thus, they coincide with $\boldsymbol{\Sigma_{MR}}$ and $\boldsymbol{\hat{\Sigma}_{MR}}$. Even in such cases, however, their formula has not been used in practice. Econometric software packages such as Stata do not estimate their asymptotic variance but estimate the standard GMM one assuming correct specification, which leads to wrong standard errors. The main contribution of this paper is to make a novel observation that 2SLS using multiple instruments under treatment effect heterogeneity is a special case of misspecified GMM of Hall and Inoue (2003). Specifically, Proposition 1 is a special case of their Theorem 2.

Since the 2SLS estimator is the linear GMM using $\left(\mathbf{Z}'\mathbf{Z}\right)^{-1}$ as a weight matrix, we may consider an alternative GMM estimator based on another weight matrix. This will lead to a different weighted average of LATEs, which may be more appealing than the conventional 2SLS estimand. Let $E[\mathbf{L}_{i}\mathbf{L}_{i}']$ be an alternative symmetric positive definite matrix where $\mathbf{L}_{i}$ is an $(l+q)\times1$ vector, and let $\left(\mathbf{L}'\mathbf{L}\right)^{-1}$ be the sample weight matrix, where $\mathbf{L}$ is an $n\times(l+q)$ matrix. The alternative GMM estimator based on the same moment condition but a different weight matrix is given by 
\begin{equation}
\boldsymbol{\tilde{\beta}} = (\mathbf{X}'\mathbf{Z}(\mathbf{L}'\mathbf{L})^{-1}\mathbf{Z}'\mathbf{X})^{-1}\mathbf{X}'\mathbf{Z}(\mathbf{L}'\mathbf{L})^{-1}\mathbf{Z}'\mathbf{Y}.
\label{2slsA}
\end{equation}
Let $\boldsymbol{\beta_{a}}$ be the probability limit of $\boldsymbol{\tilde{\beta}}$. The asymptotic distribution of $\sqrt{n}(\boldsymbol{\tilde{\beta}}-\boldsymbol{\beta_{a}})$ and a consistent variance estimator can be obtained by replacing $\mathbf{Z}_{i}\mathbf{Z}_{i}'$ with $\mathbf{L}_{i}\mathbf{L}_{i}'$, $\mathbf{Z}'\mathbf{Z}$ with $\mathbf{L}'\mathbf{L}$, and $\hat{e}_{i}$ with $\tilde{e}_{i}=Y_{i}-\mathbf{X}'_{i}\boldsymbol{\tilde{\beta}}$, whenever they appear in Propositions 1 and 2.


\begin{remark}[Using the propensity score as an instrument]
\normalfont
When $D_{i}$ is binary, Heckman and Vytlacil (2005) show that the propensity score, $P(D_{i}=1|\mathbf{W}_{i},\mathbf{Z}_{i})$, has a few desirable properties when used as an instrument given that it is correctly specified. With the same set of instruments and covariates in the 2SLS first stage, one can estimate the (nonlinear) propensity score. The resulting IV estimator would differ from the 2SLS, but both are valid based on different weighting. One may wonder if the proposed variance estimator can be used for the IV estimator in this case. Since it is not a linear GMM estimator, the proposed formula cannot be used to calculate the standard error. Instead, the theory of two-step estimators of Newey and McFadden (1994) can be applied. The formula for a logit first stage with covariates is given in the appendix. On the other hand, the 2SLS with a fully saturated first stage would yield the same point estimate because the first stage consistently estimates the propensity score. In this case, the multiple-LATEs-robust standard error can be used. 
\end{remark}
\begin{remark}[Invalid instrument]
\normalfont
The multiple-LATEs-robust variance estimator $\boldsymbol{\hat{\Sigma}_{MR}}$ is also robust to invalid instruments, i.e., instruments correlated with the error term. Consider a linear structural model $Y_{i} = \mathbf{X}_{i}'\boldsymbol{\beta_{0}} + e_{i}$
where $\mathbf{X}_{i}$ is a $(k+p)\times1$ vector of regressors. Among $k+p$ regressors, $p$ are endogeneous, i.e. $E[\mathbf{X}_{i}e_{i}]\neq0$. If a $k+q$ vector of instruments $\mathbf{Z}_{i}$ is available such that $E[\mathbf{Z}_{i}e_{i}]=0$ and $q\geq p$, then $\boldsymbol{\beta_{0}}$ can be consistently estimated by 2SLS or GMM. If any of the instruments is invalid, then $E[\mathbf{Z}_{i}e_{i}]\neq0$ and $\boldsymbol{\beta_{0}}$ may not be consistently estimated. Instead, a pseudo-true value that minimizes the corresponding GMM criterion is estimated. Since the moment condition does not hold, the model is misspecified. There are two types of misspecification: (i) fixed or global misspecification such that $E[\mathbf{Z}_{i}e_{i}] = \boldsymbol{\delta}$ where $\boldsymbol{\delta}$ is a constant vector containing at least one non-zero component, and (ii) local misspecification such that $E[\mathbf{Z}_{i}e_{i}] = n^{-r}\boldsymbol{\delta}$ for some $r> 0$. A particular choice of $r=1/2$ has beeen used to analyse the asymptotic behavior of 2SLS estimators with invalid instruments by Hahn and Hausman (2005), Bravo (2010), Berkowitz, Caner, and Fang (2008, 2012), Otsu (2011), Guggenberger (2012), and DiTraglia (2015). Under either fixed or local misspecification, $\boldsymbol{\hat{\Sigma}_{MR}}$ in Proposition 2 is consistent for the true asymptotic variance. However, the conventional variance estimator $\boldsymbol{\hat{\Sigma}_{C}}$ is inconsistent under fixed misspecification. Under local misspecification, $\boldsymbol{\hat{\Sigma}_{C}}$ is consistent but the rate of convergence is negatively affected. 
\end{remark}

\begin{remark}[Bootstrap]
\normalfont
Bootstrapping can be used to get more accurate $t$ tests and confidence intervals (CI's) based on $\boldsymbol{\hat{\beta}}$ in terms of having smaller errors in the rejection probabilities or coverage probabilities. This is called asymptotic refinements of the bootstrap. Since the model is over-identified and misspecified, and 2SLS is a special case of GMM, the misspecification-robust bootstrap for GMM of Lee (2014) achieves asymptotic refinements when applied to this case. In contrast, the conventional bootstrap methods for over-identified GMM of Hall and Horowitz (1996), Brown and Newey (2002), and Andrews (2002) assume correctly specified moment conditions. Since this implies constant treatment effects, they achieve neither asymptotic refinements nor consistency in this context. Suppose one wants to test $H_{0}: \beta_{m} = \beta_{0,m}$ or to construct a CI for $ \beta_{0,m}$ where $\beta_{0,m}$ is the $m$th element of $\boldsymbol{\beta_{0}}$. The misspecification-robust bootstrap critical values for $t$ tests and CI's are obtained from the simulated distribution of the bootstrap $t$ statistic
\begin{equation}
\nonumber T_{n}^{*} = \frac{\hat{\beta}^{*}_{m}-\hat{\beta}_{m}}{\sqrt{\hat{\Sigma}^{*}_{MR,m}/n}}
\end{equation}
where $\hat{\beta}^{*}_{m}$ and $\hat{\beta}_{m}$ are the $m$th elements of $\boldsymbol{\hat{\beta}^{*}}$ and $\boldsymbol{\hat{\beta}}$, respectively, $\hat{\Sigma}^{*}_{MR,m}$ is the $m$th diagonal element of $\boldsymbol{\hat{\Sigma}_{MR}^{*}}$, and $\boldsymbol{\hat{\beta}^{*}}$ and $\boldsymbol{\hat{\Sigma}_{MR}^{*}}$ are the bootstrap versions of $\boldsymbol{\hat{\beta}}$ and $\boldsymbol{\hat{\Sigma}_{MR}}$ based on the same formula using the bootstrap sample.
\end{remark}

\section{Weighted Averages of LATEs}
\label{weighted}

How often are researchers interested in a weighted average of LATEs? Probably more common than one might think. First of all, a discrete instrument with a binary treatment identifies a weighted average of LATEs where each LATE corresponds to a pair of two adjacent values of the instrument. This is the parameter defined in Theorem 2 of Imbens and Angrist (1994), and it is one way to average different LATEs because a discrete instrument can be written as a set of mutually exclusive binary instruments. Second, a binary instrument identifies the average causal response (ACR, Angrist and Imbens, 1995) when the treatment variable takes multiple values, e.g. years of schooling. ACR is a weighted average of LATEs for each value of the treatment variable, and is a widely accepted concept in the literature, e.g. Bleakley and Chin (2004), Lochner and Moretti (2004), Elder and Lubotsky (2009). 

On the other hand, researchers typically use 2SLS with multiple instruments to increase efficiency and because it is often unclear which instrument gives the strongest identification, assuming that the treatment effect heterogeneity is minimal, if not constant. Examples include Angrist and Chen (2011), Angrist and Evans (1998), Angrist and Krueger (1991, 1994), Angrist, Lavy, Schlosser (2010), Evans and Garthwaite (2012), Evans and Lien (2005), Siminski (2013), Stephens Jr. and Yang (2014), and Thornton (2008), among others. Even in these cases, the multiple-LATEs-robust standard errors can provide safeguards against potential violations of minimal or constant treatment effect heterogeneity. 



For the rest of this section, I replicate three well-known studies and show the multiple-LATEs-robust standard errors can be substantially different from the reported ones. The first example is Angrist and Krueger (1991), who study the returns to education. The authors avoid endogeneity of education by instrumenting it with quarter of birth (QOB). Individuals who were born at the end of the year enter school at a younger age compared with their classmates. As a result, they are required to take more compulsory schooling before they reach a legal dropout age. Angrist and Krueger estimate the following 2SLS model:
\begin{eqnarray}
\label{AKeq2}
\ln W_{i} &=& \mathbf{X}_{i}'\boldsymbol{\beta} + \sum_{c=1}^{9}Y_{ic}\psi_{c} +  E_{i} \rho+ \epsilon_{i},\\
E_{i} &=& \mathbf{X}_{i}'\boldsymbol{\pi} + \sum_{c=1}^{9}Y_{ic}\delta_{c} + \sum_{c=1}^{10}\sum_{j=1}^{3}Y_{ic}Q_{ij}\theta_{jc} + u_{i},
\label{AKeq1}
\end{eqnarray}
where $E_{i}$ is education, $X_{i}$ is a vector of covariates including a constant, $Y_{ic}$ is year of birth (YOB), $Q_{ij}$ is QOB,
and $W_{i}$ is weekly wage. If we assume that $X_{i}$ only contains a constant, then the first stage equation \eqref{AKeq1} is saturated. In this case, the 2SLS estimand is a weighted average of returns to education where averaging takes place on three different levels. First, for each level of education, it is ATE for those who would have additional schooling due to their QOB and YOB. Second, it is the ACR averaged over different levels of education which takes values from 0 to 20. Lastly, the ACR is averaged over different years of birth. By using interaction terms between QOB and YOB dummies as instruments in the first stage, it is assumed that the level of education varies with the QOB-YOB interactions, and thus is expected to increase the model fit. If the authors were interested in the returns to education for each YOB, interaction terms between YOB and education as well as YOB dummies could be included in the second stage. Instead, what is estimated in the second stage is an average of returns to education across different years of birth while controlling for the level effect of YOB, because there is no reason that a particular year, e.g. men born in 1930, is more interesting than the cohort of those born in 1930-1939. Therefore it is important to correctly calculate the standard error of the point estimates in this example. 

\begin{table}[t]
	\small
	\centering
	\begin{tabular}{lcccccc}
		\toprule
		& \multirow{2}{*}{Column} 	& \multirow{2}{*}{$\hat{\rho}$}	& \multirow{2}{*}{C}		& \multirow{2}{*}{MR}		& p-value of & \multirow{2}{*}{$F$} \\
		& & 	& 		& 		& J test (dof) 	& \\
		\midrule
		Table IV:			&	(0)		&  .0634	&	.0166	&	\textbf{.0167}	&	.3136 (2)	& 38.37  \\ [0.2em]
		Men Born 	   &   (2)	 	&	.0769	&	.0151 & 	\textbf{.0170}	&	.1661 (29)	& 4.60  \\ [0.2em]
		1920-1929, &	 (4)	 &	.1310	&	.0336	&	\textbf{.0454}	&	.5359 (27)	& 1.09 \\ [0.2em]			
		$n=247,199$ &	 (6)	 &	.0669	&	.0152	&	\textbf{.0169}	&	.2196 (29)	& 4.57  \\ [0.2em]				
		&	 (8)	 &	.1007	&	.0336	&	\textbf{.0474}	&	.3578 (27)	& 1.03  \\ [0.2em]
		\midrule
		Table V:				&	(0)		&	.1053  &	.0201	&	\textbf{.0204}	&	.1917 (2)	& 32.32  \\ [0.2em]
		Men Born    &   (2)	   &   .0891  &		.0162 & 	\textbf{.0176}	&	.6935 (29)	& 4.80  \\ [0.2em]
		1930-1939, &	 (4)	 &	.0760	&	.0292 	&	\textbf{.0359}	&	.7110 (27)	& 1.59  \\ [0.2em]				
		$n=329,509$ &	 (6)	 &	.0806	&	.0165		&	\textbf{.0178}	&	.8184 (29)	& 4.62  \\ [0.2em]				
		&	 (8)	 &	.0600	&	.0292		&	\textbf{.0349}	&	.8614 (27)	& 1.58 \\ [0.2em]		
		\midrule		
		Table VI			&	(0)		&	-.0612	&	.0259	&	\textbf{.0275}	&	.0042 (2)	& 28.85 \\ [0.2em]
		Men Born    &   (2)	   &  .0553	   &	.0138 & 	\textbf{.0166}	&	.0000 (29)	& 7.27 \\ [0.2em]
		1940-1949, &	 (4)	 &	.0948   &	.0221 	&	\textbf{.0277}		&	.0049 (27)	& 3.18\\ [0.2em]				
		$n=486,926$ &	 (6)	 &	.0393	&	.0146		&	\textbf{.0175}		&	.0000 (29)	& 6.54\\ [0.2em]				
		&	 (8)	 &	.0779 	&	.0238		&	\textbf{.0308}		&	.0033 (27)	& 2.70\\
		\bottomrule
	\end{tabular}		
	\caption{Comparison of the proposed multiple-LATEs-robust (MR) and the conventional (C) standard errors---Replication of Table IV, V, and VI in Angrist and Krueger (1991). Each column corresponds to different sets of covariates and instruments. Column (0) refers to the case that only the three quarter-of-birth dummies are used as instruments, calculated by the author.}
	\label{AK}
\end{table}

Table \ref{AK} shows replication results of Tables IV-VI in Angrist and Krueger along with the multiple-LATEs-robust standard errors (Column MR, in bold). The results for covariates are suppressed. There are a few interesting findings. First, even with large sample sizes, the two standard errors are substantially different. The conventional ones (Column C) are underestimated in all specifications. Second, large p-values do not necessarily mean that the two standard errors are similar. Third, the first stage F statistics are below the rule of thumb, 10, except for the Column (0), indicating that the instruments may be weak. One may wonder if the difference between the two standard errors are driven by the weak instruments. In the next section, I provide a simulation result that the two standard errors can be substantially different even when the F statistics are very large, although there is a weak negative relationship between the difference of the two standard errors and the magnitude of the F statistic. Finally, point estimates averaged over a large set of instruments are more robust. This case is illustrated by Table 3 Column (0) where three QOB dummies are used as only instruments with YOB dummies as covariates. Surprisingly, the return to education is estimated to be negative and significant. Further inspection reveals that it is a linear combination of three IV estimates, -0.0191 (0.0272) using only the first quarter as an instrument, -1.3167 (5.2517) using the second quarter, 0.2858 (0.1932) using the third quarter, where the numbers in parentheses are conventional IV standard errors. Apparently, an imprecisely estimated point estimate with the second QOB is the main reason for the negative point estimate in Column (0). In practice, a researcher might get around a similar problem by using a different instrument, but a better alternative is to get the 2SLS estimate based on a larger set of instruments. 

The second example is Angrist and Evans (1998) who use the sex of mother's first two children as instruments to estimate the effect of family size on mother's labor supply. The instruments \textit{two-boys} and \textit{two-girls} are based on the fact that American parents tended to have a third child when their first two children were of the same sex. Each of the instruments identifies LATE of those whose fertility was affected by their children's sex mix, and the two LATEs are not necessarily the same. For instance, a subpopulation with certain cultural background may have relatively large number of \textit{two-girls} compliers, and their ATE may be lower than that of \textit{two-boys} compliers in the population. The 2SLS model used by Angrist and Evans is
\begin{eqnarray}
\label{AE1s}
Y_{i} &=& \mathbf{X}_{i}'\boldsymbol{\beta} + M_{i}\rho +  \epsilon_{i},\\
M_{i} &=& \mathbf{X}_{i}'\boldsymbol{\pi} + TB_{i}\cdot\theta_{1} + TG_{i}\cdot\theta_{2} + u_{i},
\label{AE2s}
\end{eqnarray} 
where $M_{i}$ is an indicator for more than 2 children, $TB_{i}$ and $TG_{i}$ are indicators for first two boys and first two girls, $X_{i}$ is a vector of covariates including mother's age, age at first birth, race and Hispanic indicators, a firstborn boy indicator, and a constant, and $Y_{i}$ is an indicator for whether the respondent worked for pay in the Census year. The OLS estimate of $\rho$ is -.167, but it is argued to exaggerate the causal effect of fertility on female labor supply due to selection bias. Using the instruments one at a time, we get the IV estimates -.201 (.045) for \textit{two-boys} and -.059 (.035) for \textit{two-girls} instrument, where the numbers in parentheses are conventional IV standard errors. The estimates are quite different, and it is also difficult to compare them with the OLS estimate because the latter is for the whole population while the IV estimates are for complier subpopulations. Since the ultimate goal is to estimate the overall effect of having more than two children on mother's labor supply, one way to proceed is to calculate an average of the two IV estimates. 2SLS estimand is a particular weighted average where the weights are calculated based on the relative strength of each instrument.\footnote{In this example, 2SLS estimand is not exactly equal to a weighted average of covariate-specific LATEs, because the first stage is not fully saturated. This may weaken its causal interpretation. Nevertheless, Angrist (2001) shows that 2SLS estimates are very similar to those based on a semiparametric procedure of Abadie (2003) which allows robust causal interpretations, and Angrist and Pischke (2009) argue that this is likely to hold for other cases.} 

\begin{table}[t]
	\small
	\centering
	\begin{tabular}{ccccccc}
		\toprule
		\multirow{2}{*}{$Y_{i}$}	& \multirow{2}{*}{Estimator}  &\multirow{2}{*}{$\hat{\rho}$} &\multirow{2}{*}{C}		& \multirow{2}{*}{MR}		& p-value of &\multirow{2}{*}{$F$} \\
		& && 		& 		& J test (dof) & \\
		\midrule
		Worked for pay	& 2SLS (both)	&  -.1128	&	.0277	&	\textbf{.0277}	&	.0129 (1)&714.9\\ [0.2em]
		& IV (two-boys) & -.2011 & \multicolumn{2}{c}{.0450} & -- & 543.9\\ [0.2em]
		& IV (two-girls) & -.0591 & \multicolumn{2}{c}{.0352} & -- & 885.7\\ [0.2em]
		& OLS					& -.1666 & \multicolumn{2}{c}{.0020} & -- & --\\ [0.2em]
		\midrule
		Weeks worked & 2SLS (both)	 	&	-5.164	&	1.201 & 	\textbf{1.203}	&	.0711 (1)	& 714.9 \\ [0.2em]
		& IV (two-boys) & -7.944 & \multicolumn{2}{c}{1.950} & -- & 543.9\\ [0.2em]
		& IV (two-girls) & -3.473 & \multicolumn{2}{c}{1.527} & -- &885.7\\ [0.2em]
		& OLS					& -8.044 & \multicolumn{2}{c}{.087} & -- &--\\ [0.2em]
		\midrule
		Hours/week	 & 2SLS (both)			&	-4.613	&	1.008	&	\textbf{1.010}	&	.0492 (1)	& 714.9\\ [0.2em]
		& IV (two-boys) & -7.159 & \multicolumn{2}{c}{1.644} & -- & 543.9\\ [0.2em]
		& IV (two-girls) & -3.065 & \multicolumn{2}{c}{1.279} & -- &885.7\\ [0.2em]
		& OLS					& -6.021 & \multicolumn{2}{c}{.074} & -- &--\\ [0.2em]
		\midrule
		Labor income	& 2SLS (both)		&	-1321.2	&	566.4	&	\textbf{566.4}	&	.7025 (1)	& 714.9\\ [0.2em]
		& IV (two-boys) & -1597.8 & \multicolumn{2}{c}{914.7}  & -- & 543.9\\ [0.2em]
		& IV (two-girls) & -1153.0 & \multicolumn{2}{c}{721.3} & -- &885.7\\ [0.2em]
		& OLS					& -3165.4 & \multicolumn{2}{c}{40.6} & -- &--\\ [0.2em]
		\bottomrule			
	\end{tabular}
	\caption{Comparison of the proposed multiple-LATEs-robust (MR) and the conventional (C) standard errors---Replication of Table 7 Columns (4) and (6) in Angrist and Evans (1998). The number of observations is $n=254,654$. The IV estimators using either the two-boys or two-girls instrument are calculated by the author.}
	\label{AE}
\end{table}

Table \ref{AE} shows replication results of Table 7 in Angrist and Evans (1998). First of all, the 2SLS estimates are smaller in magnitude than the OLS estimates, providing evidence that the OLS overestimates the true effect. Second, unlike the replication of Angrist and Krueger (1991), the two standard errors are almost the same, even when the p-values are quite small. Since they are similar, there is a sizable gain in precision by combining the two instruments compared with using a single instrument even when we use the multiple-LATEs-robust standard errors. Note that standard errors converge in probability to zero. Thus, the similarity of two standard errors in Table \ref{AE} should not be misinterpreted as the similarity of the asymptotic variances. Lastly, the 2SLS point estimates are weighted averages of the two IV estimates, where the weight for \textit{two-boys} instrument is .38. Since the weight is completely determined by the first stage, the same weight is used across different dependent variables in the second stage. In this example, \textit{two-boys} instrument receives less weight because the first-stage coefficient is smaller, which implies that the absolute size of the compliers is smaller than that of \textit{two-girls} instrument. The multiple-LATES-robust standard error can be computed not only for 2SLS but also for any other weighted averages of LATEs, as long as they can be written as a GMM estimator as in \eqref{2slsA}.

The last example is Thornton (2008), who studies the impact of learning HIV status on purchasing condoms in rural Malawi using randomly assigned monetary incentives and distance from results centers as instruments. The dependent variable $Y_{ij}$ is an indicator for condom purchase, reported buying condoms, or reported having sex at the follow up survey, or the number of condoms bought for person $i$ in village $j$. The endogenous variables are $G_{ij}$ and $G_{ij}\times HIV_{ij}$, where $G_{ij}$ indicates knowledge of HIV status and $HIV_{ij}$ indicates an HIV-positive diagnosis. In particular, the interaction term $G_{ij}\times HIV_{ij}$ is included in the main equation to investigate the differential effect of receiving a positive HIV diagnosis. Both endogenous variables are instrumented with the same set of variables. The 2SLS model used by Thornton is
\begin{eqnarray}
Y_{ij} &=& \mathbf{X}_{ij}'\boldsymbol{\beta} + G_{ij}\cdot\rho_{1} + (G_{ij}\times HIV_{ij})\cdot\rho_{2}+\epsilon_{ij},\\
G_{ij} &=& \mathbf{X}_{ij}'\boldsymbol{\pi} + \sum_{c=1}^{5}Z_{ijc}\theta_{1c} + \sum_{c=1}^{5}Z_{ijc}Male_{ij}\theta_{2c}\\
&& + \sum_{c=1}^{5}Z_{ijc}HIV_{ij}\theta_{3c} + \sum_{c=1}^{5}Z_{ijc}Male_{ij}HIV_{ij}\theta_{4c}+ u_{ij},
\end{eqnarray}
where $Z_{ijc}$ for $c=1,...,5$ are being offered any incentive, the amount of the incentive, the amount of the incentive squared, the distance from the HIV result center, and distance-squared, and $X_{i}$ is a vector of covariates including an indicator for male, $Male_{ij}$, as well as age, age-squared, a district dummy, $HIV_{ij}$, and a constant. The first stage for $G_{ij}\times HIV_{ij}$ has the same set of regressors. To account for the differential effects of gender and HIV-positive diagnosis on getting the test results, the interaction terms are also used as instruments. The author notes that the resulting estimates of $\rho_{1}$ and $\rho_{2}$ are weighted averages of LATEs, but argues that differences in LATEs across instruments may be minimal, which justifies the use of the conventional standard errors. 

Table \ref{AT} show replication results of Table 7 in Thornton (2008). The two standard errors are quite different, especially for $\hat{\rho}_{2}$. The point estimate of $\rho_{2}$ with $Y_{ij}=$\textit{Number of condoms bought} is significantly different from zero at 5\% level using the conventional standard error, but it is no longer significant even at 10\% level when the multiple-LATEs-robust standard error is used. Thus, even though the treatment effects heterogeneity is assumed to be minimal, one should report the multiple-LATEs-robust standard errors to make correct inferences. The J test is conducted using the cluster-robust variance matrix estimator and the p-values are under any reasonable nominal size. The column under $CD$ shows the Cragg-Donald statistic for multiple endogenous variables (Cragg and Donald, 1993), which can be used to measure the weak instrument problem using the Stock and Yogo (2005) weak instrument critical values under iid and homoskedasticity assumptions.

\begin{table}[t]
	\small
	\centering
	\begin{tabular}{ccccccccc}
		\toprule
		\multirow{2}{*}{$Y_{ij}$}	&  &\multirow{2}{*}{$\hat{\rho}$} &\multicolumn{2}{c}{cluster-robust} & \multicolumn{2}{c}{no cluster} & p-value of & \multirow{2}{*}{CD}\\
		& && 	C	& 	MR	& C	& 	MR	&J test (dof) & \\
		\midrule
		Bought	& $\rho_{1}$	&  -.0687	&	.0620	&	\textbf{.0639}	&	.0630 & \textbf{.0638} & \multirow{2}{*}{.0212(18)} & \multirow{2}{*}{11.77} \\ [0.2em]
		condoms & $\rho_{2}$ & .2482 & .1690 & \textbf{.2738} & .1960 & \textbf{.2591} & &\\ [0.2em]
		\midrule
		Number of  & $\rho_{1}$	 	&	-.3029	&	.2854 & 	\textbf{.2961}	&	.3093	& \textbf{.3129} & \multirow{2}{*}{.0246 (18)} &\multirow{2}{*}{11.77}\\ [0.2em]
		condoms bought & $\rho_{2}$ & 1.689 & .7839 & \textbf{1.0536} & .8030 & \textbf{.9838} & &\\ [0.2em]
		\midrule
		Reported	 & $\rho_{1}$			&	.0171	&	.0500	&	\textbf{.0493}	&	.0474	& \textbf{.0488} & \multirow{2}{*}{.0023 (18)}& \multirow{2}{*}{11.77}\\ [0.2em]
		buying condoms & $\rho_{2}$ & -.0268 & .0923 & \textbf{.1411} & .0988 & \textbf{.1773} & &\\ [0.2em]
		\midrule
		Reported having	&$\rho_{1}$	&	-.0043	&	.0598	&	\textbf{.0627}	& .0736 & \textbf{.0747}	& \multirow{2}{*}{.0200 (18)}& \multirow{2}{*}{11.77}\\ [0.2em]
		sex at follow-up& $\rho_{2}$ & -.0795 & .2288  & \textbf{.2736} & .2380 & \textbf{.2986} & &\\ [0.2em]
		\bottomrule			
	\end{tabular}
	\caption{Comparison of the proposed multiple-LATEs-robust (MR) and the conventional (C) standard errors---Replication of Table 7 Columns (2), (4), (6), and (8) in Thornton (2008). The number of observations is $n=1,008$. Standard errors clustered by village (for 57 villages). The formula for the cluster-and-multiple-LATEs-robust standard error is given in the appendix.}
	\label{AT}
\end{table}


\section{Simulation}
\label{simulation}

This section provides simulation evidence to answer three questions: (i) whether the multiple-LATEs-robust standard error correctly approximates the true standard deviation of the 2SLS estimator, (ii) whether the difference between the two standard errors is driven by the weak instruments, and (iii) whether there is a relationship between the difference in the magnitude of the two standard errors and the p-value of the J test. 

I generate a random subsample without replacement from the 1980 Census Public Use Micro Samples dataset of Angrist and Evans (1998) that contains information on $254,654$ married women. The 2SLS model is the same as  \eqref{AE1s} and \eqref{AE2s}, except that an indicator for multiple second birth is also added as an instrument. Angrist and Evans show evidence that this instrument identifies LATE different from that of \textit{two-boys} or \textit{two-girls} instruments, and give further analysis on the issue. For simulation, I simply use three instruments together without introducing additional parameters that partly adjust for the difference in LATEs, so that the underlying moment condition model is over-identified and potentially misspecified.

\begin{table}[t]
	\small
	\centering
	\begin{tabular}{lcrrccc}
		\toprule
		\multirow{2}{*}{$Y_{i}$}	& \multirow{2}{*}{$n$} & \multicolumn{2}{c}{$\hat{\rho}$}	 & C	& MR & $J$ \\
		\cmidrule{3-7}
		& & mean	& s.d. & mean		& mean		 & Rej 5\% \\
		\midrule
		\multirow{3}{*}{Worked for pay} & 1,000 & -.1005 & \textbf{.2331} & .2241 &	\textbf{.2351}		 &.0564\\ 
															  & 2,000 &  -.0932 & \textbf{.1657} & .1610 &	\textbf{.1647}		 &.0484\\ 
																& 5,000 & - .0940 & \textbf{.1028} & .1027 & \textbf{.1036} & .0506\\
		\midrule
		\multirow{3}{*}{Weeks worked} & 1,000  &  -4.780 & \textbf{9.861} & 9.469 &	\textbf{9.939}	&	.0587\\ 
															  & 2,000  &  -4.477 & \textbf{7.026} & 6.823 &	\textbf{6.979}	&	.0529\\ 
																& 5,000  & -4.536  & \textbf{4.405} & 4.351 & \textbf{4.389} & .0492\\
		\midrule
		\multirow{3}{*}{Hours/week} 		& 1,000  &  -3.943 & \textbf{8.263} & 7.904 &	\textbf{8.298}	&	.0608\\ 
																& 2,000  &  -3.762 & \textbf{5.867} & 5.677 &	\textbf{5.810}	&	.0565\\ 
																& 5,000  & -3.875  & \textbf{3.630} & 3.614 & \textbf{3.648}& .0532\\
		\midrule
		\multirow{3}{*}{Labor income}  & 1,000  &  -1527.6 & \textbf{4564.8} & 4168.0 &	\textbf{4398.1}	&	.0883\\ 
															& 2,000  &  -1397.3 & \textbf{3167.0} & 3051.7 &	\textbf{3127.6}	&	.0772\\ 
															& 5,000  & -1446.5  & \textbf{1991.7} & 1959.3 & \textbf{1977.0}& .0599\\
		\bottomrule			
	\end{tabular}
	\caption{The mean, standard deviation and standard errors of the 2SLS estimator based on a subsample of Angrist and Evans (1998).}
	\label{T1}
\end{table}

Table \ref{T1} shows the mean and standard deviation of the 2SLS estimator, and the means of the multiple-LATEs-robust (MR) and the conventional standard errors (C) for different dependent variables based on 10,000 replications for the sample sizes 1,000, 2,000, and 5,000. The proportion of the heteroskedasticity-robust first-stage F statistics smaller than the rule of thumb 10 is 0.25\% for $n=1,000$, and zero for the other sample sizes. Thus, there is little concern for weak instruments. In an unreported simulation result, the multiple-LATEs-robust standard error tends to approximate the standard deviation of the 2SLS estimator well with marginal F statistics under 10. Across specifications, the conventional standard error based on $\boldsymbol{\hat{\Sigma}_{C}}$ underestimates the standard deviation. Therefore, inferences based on the conventional standard errors can be misleading. In contrast, the multiple-LATEs-robust standard error estimates the standard deviation more accurately. The last column shows the actual rejection probability of the J test at 5\% significance level. Since the p-values for the full sample ($n=254,654$) are .0345, .1599, .0971, and .9158 for each dependent variable, respectively, the result shows that the J test exhibits very low power, at least for $Y_{i}=\textit{Worked for pay}$ and $Y_{i}=\textit{Hours per week}$.

\begin{figure}[t]
	\centering
	\begin{subfigure}[t]{0.48\textwidth}
		\centering
		\includegraphics[width=77mm]{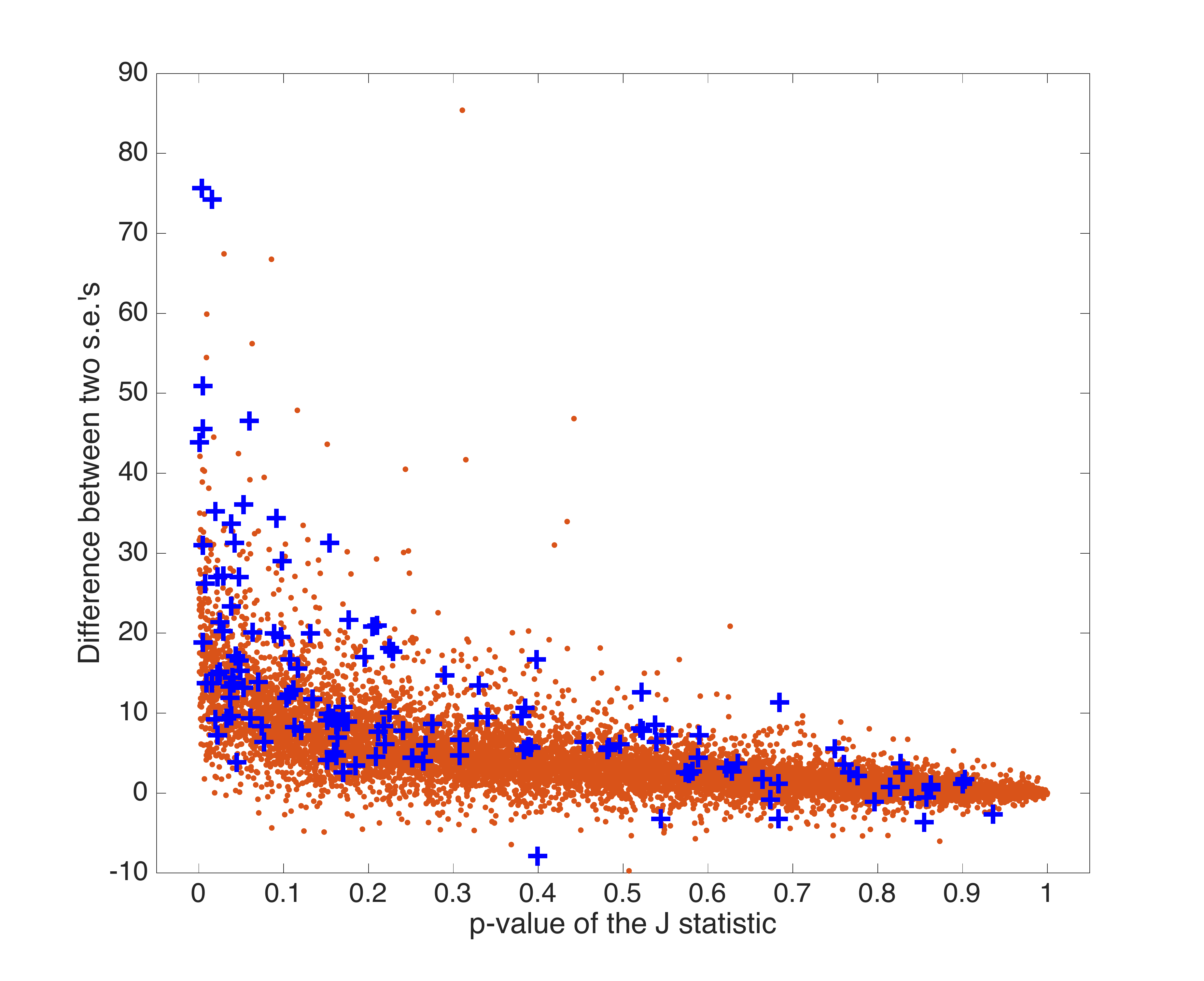}		
		\caption{J statistic, n=1,000}
	\end{subfigure}%
~
	\begin{subfigure}[t]{0.48\textwidth}
		\centering
		\includegraphics[width=77mm]{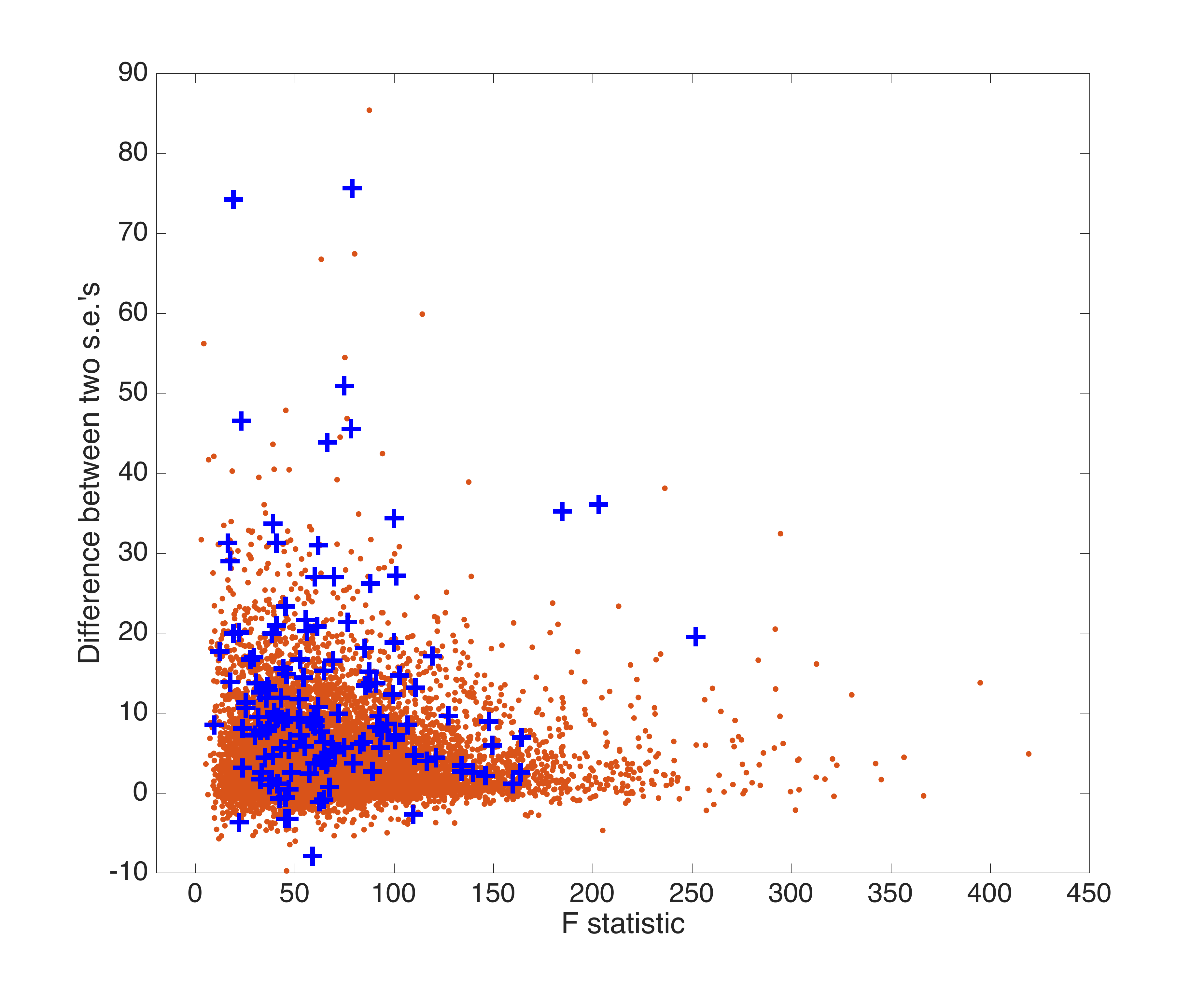}		
		\caption{F statistic, n=1,000}
	\end{subfigure}

	\begin{subfigure}[t]{0.48\textwidth}
		\centering
	 	\includegraphics[width=77mm]{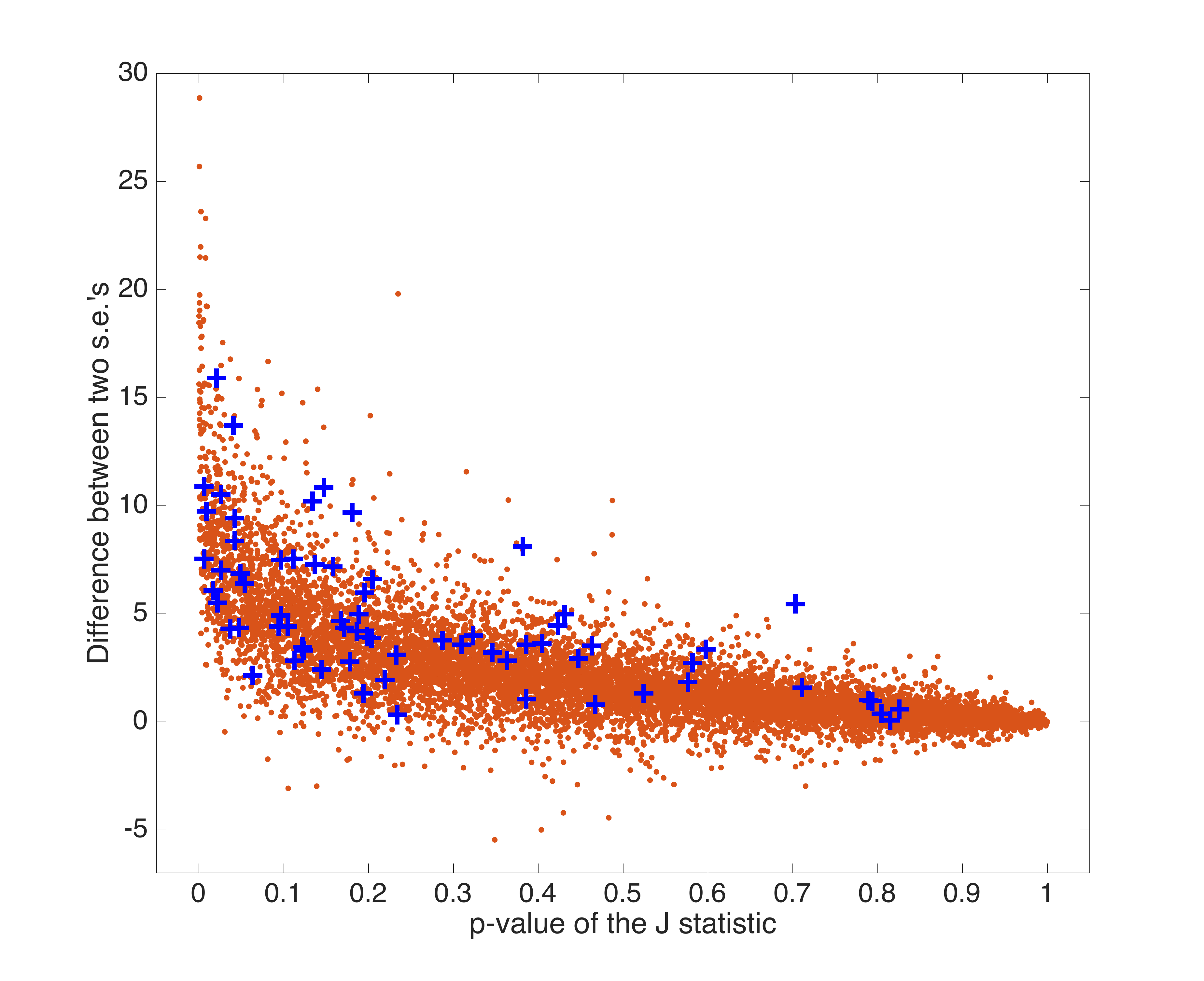}		
		\caption{J statistic, n=2,000}
	\end{subfigure}
~
	\begin{subfigure}[t]{0.48\textwidth}
		\centering
	 	\includegraphics[width=77mm]{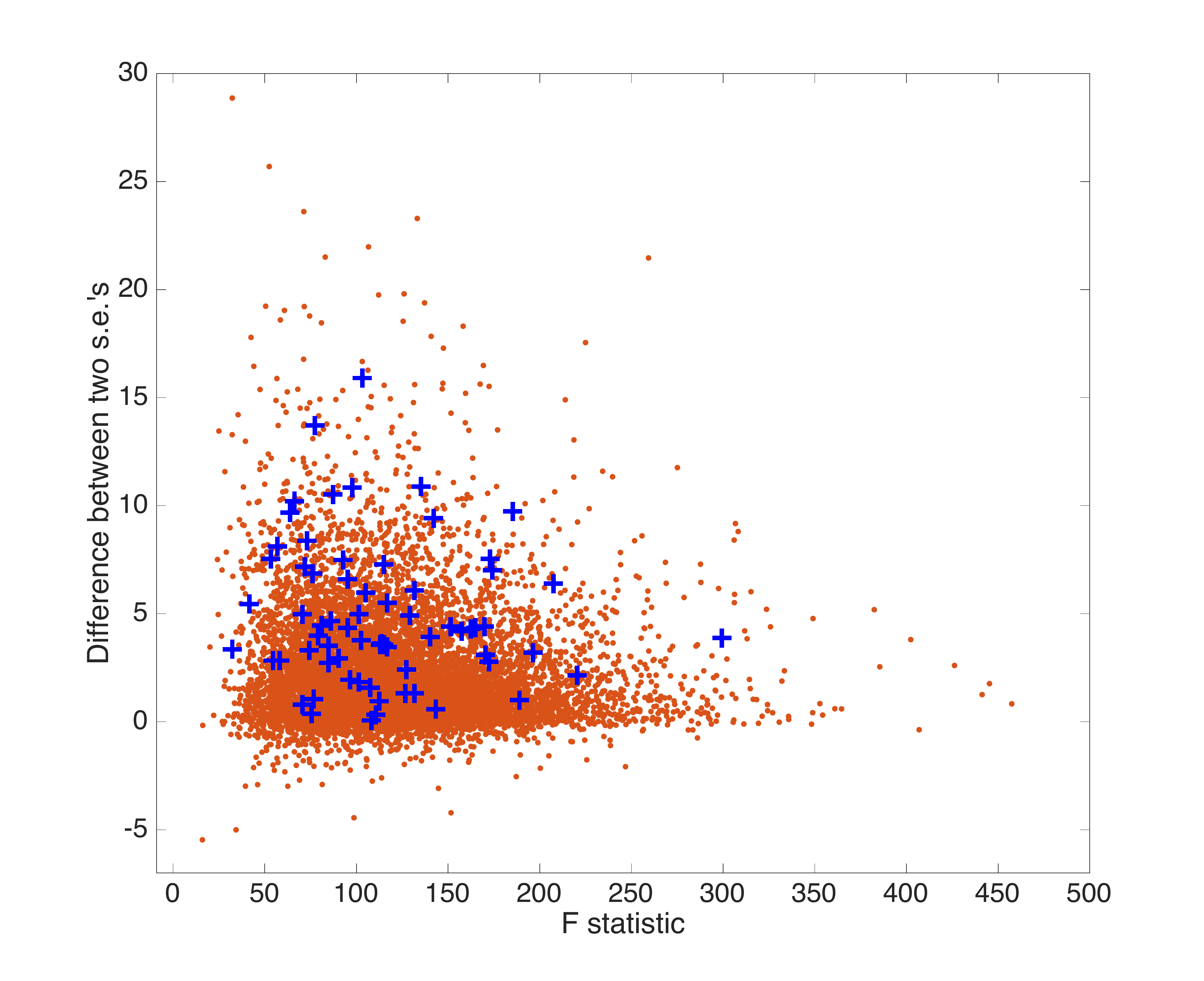}		
		\caption{F statistic, n=2,000}
	\end{subfigure}
 	\begin{subfigure}[t]{0.48\textwidth}
 		\centering
	 	\includegraphics[width=77mm]{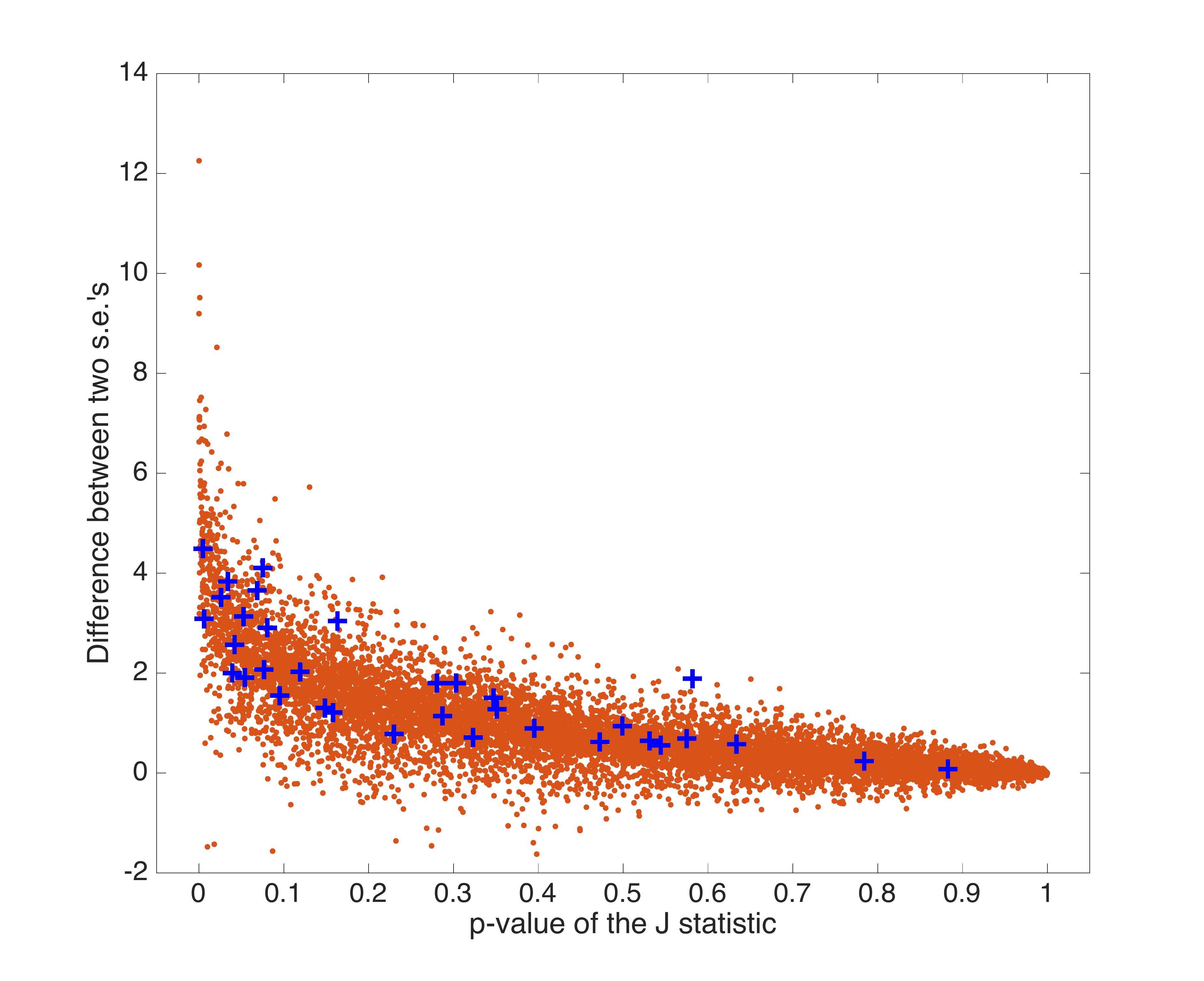}		
 		\caption{J statistic, n=5,000}
 	\end{subfigure}
 	~
 	\begin{subfigure}[t]{0.48\textwidth}
 		\centering
	 	\includegraphics[width=77mm]{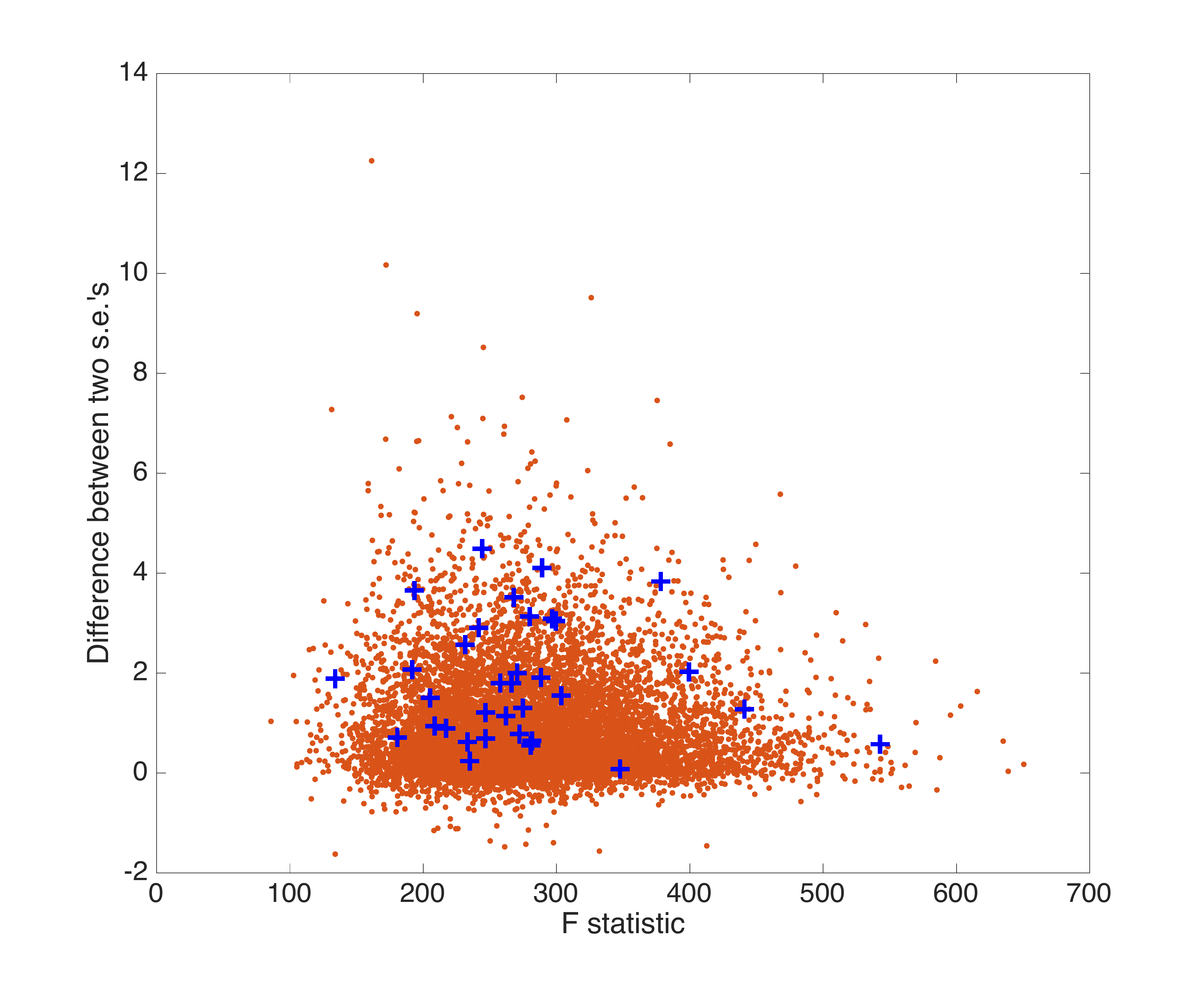}		
 		\caption{F statistic, n=5,000}
 	\end{subfigure}
\caption{Scatterplots of p-values of the J test and F statistics, and percentage difference between two standard errors, $s.e._{\boldsymbol{\hat{\Sigma}_{MR}}}$ and $s.e._{\boldsymbol{\hat{\Sigma}_{C}}}$ }
 \label{fig1}
\end{figure}

Figure \ref{fig1} shows the relationships between the percentage difference between the two standard errors (the mutiple-LATEs robust standard error minus the conventional standard error divided by the average of the two) and the p-value of the J test, and the F statistic, respectively, for different sample sizes when $Y_{i}=\textit{Worked for pay}$. The results are similar for other dependent variables, and thus not reported. The multiple-LATEs-robust standard error is likely to differ much from the conventional one when the p-value of the J test or the F statistic is relatively small, though the negative correlation is quite weak for the F statistic. Points with a $+$ marker indicate that the conclusion of the $t$ tests changes with the correct standard error where the null hypothesis is that the coefficient equals to zero, i.e. statistically significant results may not be significant anymore (or the other way around). In particular, this happens regardless of the p-value of the J test or the magnitude of the F statistic.

In sum, the simulation result shows that the multiple-LATEs-robust standard error can be very different from the conventional one even with large F statistics and p-values of the J test. Furthermore, the conclusion of the empirical study may change if the multiple-LATEs-robust standard error is used. 

\section{Conclusion}
\label{conclusion}
Two-stage least squares (2SLS) estimators are widely used in practice. When heterogeneity is present in treatment effects, 2SLS point estimates can be interpreted as weighted averages of the local average treatment effects (LATE). I show that the conventional standard errors, typically generated by econometric software packages such as Stata, are incorrect in this case. The over-identifying restrictions test is often used to test the presence of heterogeneity, but it is not useful in this context because it can also reject due to invalid instruments. I provide a simple standard error formula for 2SLS which is correct regardless of whether there are multiple LATEs or not. In addition, this standard error is robust to invalid instruments, and can be used for bootstrapping to achieve asymptotic refinements under treatment effect heterogeneity. I recommend practitioners to always use the proposed multiple-LATEs-robust standard error for 2SLS.

\appendix
\section*{Appendix}

\section{Proofs of Propositions}
\subsection*{Proposition 1}
\begin{proof}
Let $\boldsymbol{e} \equiv (e_{1},...,e_{n})'$ be an $n\times1$ vector where $e_{i} \equiv Y_{i}-\mathbf{X}_{i}'\boldsymbol{\beta_{0}}$. I claim that evaluated at $\boldsymbol{\beta_{0}}$, the moment condition does not hold:
\begin{equation}
E[\mathbf{Z}_{i}(Y_{i}-\mathbf{X}_{i}'\boldsymbol{\beta_{0}})]\equiv E[\mathbf{Z}_{i}e_{i}]\neq0,
\label{mis}
\end{equation}
if there is more than one LATE. This can be shown by the following argument. For simplicity, assume that we have two instruments, $Z_{i}^{1}$ and $Z_{i}^{2}$, such that each instrument satisfies regularity conditions for identifying the instrument-specific LATE. Let $\rho_{j}$ be the LATE with respect to $Z_{i}^{j}$ and $\boldsymbol{\beta}_{j}\equiv (\boldsymbol{\gamma_{j}}',\rho_{j})'$ be the parameter vector for $j=1,2$. By assumption, $\boldsymbol{\beta_{1}}\neq\boldsymbol{\beta_{2}}$. If we use each instrument one at a time, $E[Z_{i}^{1}(Y_{i}-\mathbf{X}_{i}'\boldsymbol{\beta_{1}})]=E[Z_{i}^{2}(Y_{i}-\mathbf{X}_{i}'\boldsymbol{\beta_{2}})]=0$. Now assume $E[\mathbf{Z}_{i}(Y_{i}-\mathbf{X}_{i}'\boldsymbol{\beta_{0}})]=0$ holds. Then $E[Z_{i}^{1}(Y_{i}-\mathbf{X}_{i}'\boldsymbol{\beta_{0}})]=E[Z_{i}^{2}(Y_{i}-\mathbf{X}_{i}'\boldsymbol{\beta_{0}})]=0$, but this implies $\boldsymbol{\beta_{0}}=\boldsymbol{\beta_{1}}=\boldsymbol{\beta_{2}}$. This contradicts the assumption. Thus, \eqref{mis} holds.

From the GMM FOC, we substitute $\mathbf{X}\boldsymbol{\beta_{0}} + \boldsymbol{e}$ for $\mathbf{Y}$, rearrange terms, and multiply $\sqrt{n}$ to have
\begin{eqnarray}
\label{expan}
\sqrt{n}(\boldsymbol{\hat{\beta}}-\boldsymbol{\beta_{0}}) &=& (\mathbf{X}'\mathbf{Z}(\mathbf{Z}'\mathbf{Z})^{-1}\mathbf{Z}'\mathbf{X})^{-1}\mathbf{X}'\mathbf{Z}(\mathbf{Z}'\mathbf{Z})^{-1}\sqrt{n}\mathbf{Z}'\boldsymbol{e},\\
\nonumber &=& \left(\frac{1}{n}\mathbf{X}'\mathbf{Z}\left(\frac{1}{n}\mathbf{Z}'\mathbf{Z}\right)^{-1}\frac{1}{n}\mathbf{Z}'\mathbf{X}\right)^{-1}\times\\
\nonumber && \left\{\frac{1}{n}\mathbf{X}'\mathbf{Z}\left(\frac{1}{n}\mathbf{Z}'\mathbf{Z}\right)^{-1}\sqrt{n}\left(\frac{1}{n}\mathbf{Z}'\boldsymbol{e} -E[\mathbf{Z}_{i}e_{i}]\right)\right.\\
\nonumber && + \sqrt{n}\left(\frac{1}{n}\mathbf{X}'\mathbf{Z}- E[\mathbf{X}_{i}\mathbf{Z}_{i}']\right)\left(\frac{1}{n}\mathbf{Z}'\mathbf{Z}\right)^{-1}E[\mathbf{Z}_{i}e_{i}] \\
\nonumber && + \left. E[\mathbf{X}_{i}\mathbf{Z}_{i}']\sqrt{n}\left(\left(\frac{1}{n}\mathbf{Z}'\mathbf{Z}\right)^{-1}-\left(E[\mathbf{Z}_{i}\mathbf{Z}_{i}']\right)^{-1}\right)E[\mathbf{Z}_{i}e_{i}]\right\}.
\end{eqnarray}
The second equality holds because the population GMM FOC holds regardless of misspecification, i.e., $0=E[\mathbf{X}_{i}\mathbf{Z}_{i}']E[\mathbf{Z}_{i}\mathbf{Z}_{i}]^{-1}E[\mathbf{Z}_{i}e_{i}]$. The expression \eqref{expan}  is different from the standard one because $E[\mathbf{Z}_{i}e_{i}]\neq0$. As a result, the asymptotic variance matrix of $\sqrt{n}(\boldsymbol{\hat{\beta}}-\boldsymbol{\beta_{0}})$ includes additional terms, which are assumed to be zero in the standard asymptotic variance matrix of 2SLS. We use the fact that 
\begin{equation}
\left(\frac{1}{n}\mathbf{Z}'\mathbf{Z}\right)^{-1}-E[\mathbf{Z}_{i}\mathbf{Z}_{i}']^{-1} = \left(E[\mathbf{Z}_{i}\mathbf{Z}_{i}']\right)^{-1}\left(E[\mathbf{Z}_{i}\mathbf{Z}_{i}']-\frac{1}{n}\mathbf{Z}'\mathbf{Z}\right)\left(\frac{1}{n}\mathbf{Z}'\mathbf{Z}\right)^{-1},
\end{equation}
and take the limit of the right-hand-side of \eqref{expan}. By the weak law of large numbers (WLLN), the continuous mapping theorem (CMT), and the central limit theorem (CLT),
\begin{equation}
\sqrt{n}(\boldsymbol{\hat{\beta}}-\boldsymbol{\beta_{0}}) \overset{d}{\rightarrow}\textbf{H}^{-1}\cdot N(\mathbf{0},\boldsymbol{\Omega}).
\end{equation}
\end{proof}

\subsection*{Proposition 2}
\begin{proof}
Since $\boldsymbol{\hat{\beta}}$ is consistent for $\boldsymbol{\beta_{0}}$, by WLLN and CMT, $n^{-1}\sum_{i}\boldsymbol{\hat{\psi}}_{i}\boldsymbol{\hat{\psi}}_{i}'$ is consistent for $\boldsymbol{\Omega}$. By using WLLN and CMT again, $\boldsymbol{\hat{\Sigma}_{MR}}$ is consistent for $\textbf{H}^{-1}\boldsymbol{\Omega} \textbf{H}^{-1}$.
\end{proof}

\section{Asymptotic variance when the logit model is used for the first stage}
Assume the logit model for the propensity score:
\begin{equation}
P(D_{i}=1|\textbf{W}_{i},\textbf{Z}_{i}) = \frac{1}{1+\exp(-\textbf{W}_{i}'\boldsymbol{\delta}_{0}-\textbf{Z}_{i}'\boldsymbol{\pi}_{0})}.
\end{equation}
Then the log-likelihood function is
\begin{equation}
\mathcal{L}(\boldsymbol{\delta},\boldsymbol{\pi}) = -\sum_{i=1}^{n}(1-D_{i})(\textbf{W}_{i}'\boldsymbol{\delta}+\textbf{Z}_{i}'\boldsymbol{\pi})-\sum_{i=1}^{n}\ln\left(1+\exp(-\textbf{W}_{i}'\boldsymbol{\delta}-\textbf{Z}_{i}'\boldsymbol{\pi})\right).
\end{equation}
The FOC of the first stage is
\begin{equation}
0=n^{-1}\sum_{i=1}^{n}\left(\begin{array}{c}
\textbf{W}_{i} \\ \textbf{Z}_{i}
\end{array}\right)\hat{u}_{i}
\end{equation}
where
\begin{equation}
u_{i}(\boldsymbol{\delta},\boldsymbol{\pi}) = -(1-D_{i}) + \frac{\exp(-\textbf{W}_{i}'\boldsymbol{\delta}-\textbf{Z}_{i}'\boldsymbol{\pi})}{1+\exp(-\textbf{W}_{i}'\boldsymbol{\delta}-\textbf{Z}_{i}'\boldsymbol{\pi})}
\end{equation}
and $\hat{u}_{i}=u_{i}(\boldsymbol{\hat{\delta}},\boldsymbol{\hat{\pi}})$. For the second stage, the FOC is
\begin{equation}
0=n^{-1}\sum_{i=1}^{n}\left(\begin{array}{c}
\textbf{W}_{i} \\
(1+\exp(-\textbf{W}_{i}'\boldsymbol{\hat{\delta}}-\textbf{Z}_{i}'\boldsymbol{\hat{\pi}}))^{-1}
\end{array}\right)\hat{e}_{i} 
\end{equation}
where $e_{i}(\boldsymbol{\gamma},\rho) = Y_{i}-\textbf{W}_{i}'\boldsymbol{\gamma}-D_{i}'\rho$ and $\hat{e}_{i}=e_{i}(\boldsymbol{\hat{\gamma}},\boldsymbol{\hat{\rho}})$. Now consider a stacked moment function
\begin{equation}
h_{i}(\boldsymbol{\beta}) = \left(\begin{array}{c}
\textbf{W}_{i}u_{i}(\boldsymbol{\delta},\boldsymbol{\pi}) \\ \textbf{Z}_{i}u_{i}(\boldsymbol{\delta},\boldsymbol{\pi}) \\  \textbf{W}_{i}e_{i}(\boldsymbol{\gamma},\rho) \\ (1+\exp(-\textbf{W}_{i}'\boldsymbol{\delta}-\textbf{Z}_{i}'\boldsymbol{\pi}))^{-1}e_{i}(\boldsymbol{\gamma},\rho)
\end{array}\right),
\end{equation}
where $\boldsymbol{\beta} = (\boldsymbol{\delta},\boldsymbol{\pi},\boldsymbol{\gamma},\rho)'$. This forms a just-identified moment condition model. Let $\boldsymbol{\hat{\beta}} = (\boldsymbol{\hat{\delta}},\boldsymbol{\hat{\pi}},\boldsymbol{\hat{\gamma}},\hat{\rho})'$ and $\boldsymbol{\beta}_{0} = (\boldsymbol{\delta}_{0},\boldsymbol{\pi}_{0},\boldsymbol{\gamma}_{0},\rho_{0})'$ be the probability limit. Using standard asymptotic theory for just-identified GMM, the asymptotic distribution of $\sqrt{n}(\boldsymbol{\hat{\beta}}-\boldsymbol{\beta}_{0})$ is $N(0,\textbf{V})$ where $\textbf{V} = \boldsymbol{\Gamma}^{-1}\boldsymbol{\Delta}(\boldsymbol{\Gamma}')^{-1}$, $\boldsymbol{\Gamma} = E(\partial/\partial\beta')h_{i}(\boldsymbol{\beta}_{0})$, and $\boldsymbol{\Delta} = E[h_{i}(\boldsymbol{\beta}_{0})h_{i}(\boldsymbol{\beta}_{0})']$. A consistent estimator of $\textbf{V}$ can be obtained by replacing the population moments with the sample moments: $\hat{\textbf{V}} = \boldsymbol{\hat{\Gamma}}^{-1}\boldsymbol{\hat{\Delta}}(\boldsymbol{\hat{\Gamma}}')^{-1}$ where $\boldsymbol{\hat{\Gamma}} = n^{-1}\sum_{i=1}^{n}(\partial/\partial\boldsymbol{\beta}')h_{i}(\boldsymbol{\hat{\beta}})$ and $\boldsymbol{\hat{\Delta}} = n^{-1}\sum_{i=1}^{n}h_{i}(\boldsymbol{\hat{\beta}})h_{i}(\boldsymbol{\hat{\beta}})'$. The correct standard errors for $\boldsymbol{\hat{\beta}}$ can be obtained by taking the square roots of the diagonal elements of $\hat{\textbf{V}}$ divided by $n$.

\section{Cluster-and-Multiple-LATEs-Robust Variance Estimator}
The multiple-LATEs-robust variance estimator for 2SLS given in \eqref{SigmaMR} assumes iid observations, but it can be easily generalized to a setting with independent clusters where observations within a cluster are dependent. Assume there are $G$ clusters where the size of each cluster may differ. Using the same notation given in the main text, the cluster-and-multiple-LATEs-robust variance estimator is
\begin{equation}
\label{SigmaMRcluster}
\boldsymbol{\hat{\Sigma}_{CMR}} = \hat{\mathbf{H}}^{-1}\left(\frac{1}{n}\sum_{i=1}^{n}\sum_{j=1}^{n}\mathbf{1}(i,j\in\text{same cluster})\boldsymbol{\hat{\psi}}_{i}\boldsymbol{\hat{\psi}}_{j}'\right)\hat{\mathbf{H}}^{-1}.
\end{equation}
Most built-in Stata commands use finite-sample modifications by using $\sqrt{c}\hat{e}_{i}$ instead of $\hat{e}_{i}$ where $c=\frac{G}{G-1}\cdot\frac{n-1}{n-k}$ (Cameron and Miller, 2015).

\small
\section*{References}
\begin{enumerate} [leftmargin=0 in]
\item[] Abadie, A. (2003). Semiparametric instrumental variable estimation of treatment response models. Journal of Econometrics, 113(2), 231-263.
\item[] Andrews, D. W. (2002). Higher-order improvements of a computationally attractive k-step bootstrap for extremum estimators. Econometrica, 70(1), 119-162.
\item[] Angrist, J. D. (2001). Estimation of limited dependent variable models with dummy endogenous regressors. Journal of Business and Economic Statistics, 19(1).
\item[] Angrist, J. D. and Chen, S. H. (2011). Schooling and the Vietnam-Era Gl Bill: Evidence from the Draft Lottery. American Economic Journal: Applied Economics, 96-118.
\item[] Angrist, J. D. and Evans, W. N. (1998). Children and their parents' labor supply: Evidence from exogenous variation in family size. Amerian Economic Review 88(3), 450-477.
\item[] Angrist, I. D. and Fernandez-Val, I. (2013). ExtrapoLATE-ing: External Validity and. In Advances in Economics and Econometrics: Tenth World Congress (Vol. 3, p. 401). Cambridge University Press.
\item[] Angrist, J. D. and Imbens, G. W. (1995). Two-stage least squares estimation of average causal effects in models with variable treatment intensity. Journal of the American statistical Association, 90(430), 431-442.
\item[] Angrist, J. D., Imbens, G. W., and Rubin, D. B. (1996). Identification of causal effects using instrumental variables. Journal of the American statistical Association, 91(434), 444-455.
\item[] Angrist, J. D. and Krueger, A. B. (1991). Does compulsory school attendance affect schooling and earnings?. Quarterly Journal of Economics, 106(4), 979-1014.
\item[] Angrist, J. D. and Krueger, A. B. (1994). Why do World War II veterans earn more than nonveterans?. Journal of Labor Economics, 12(1), 74-97.
\item[] Angrist, J., Lavy, V., and Schlosser, A. (2010). Multiple experiments for the causal link between the quantity and quality of children. Journal of Labor Economics, 28(4), 773-824.
\item[] Angrist, J. and Pischke, J. S. (2009). Mostly Harmless Econometrics: An Empiricist’s Companion. Princeton University Press.
\item[] Berkowitz, D., Caner, M., and Fang, Y. (2008). Are nearly exogenous instruments reliable? Economics Letters 101 (1), 20-23.
\item[] Berkowitz, D., Caner, M., and Fang, Y. (2012). The validity of instruments revisited. Journal of Econometrics 166 (2), 255-266.
\item[] Bleakley, H. and Chin, A. (2004). Language skills and earnings: Evidence from childhood immigrants. Review of Economics and Statistics, 86(2), 481-496.
\item[] Bravo, F. (2010). Efficient M-estimators with auxiliary information. Journal of Statistical Planning and Inference 140 (11), 3326-3342.
\item[] Brown, B. W. and Newey, W. K. (2002). Generalized method of moments, efficient bootstrapping, and improved inference. Journal of Business and Economic Statistics, 20(4), 507-517.
\item[] Cameron, A. C. and Miller, D. L. (2015). A practitioner's guide to cluster-robust inference. Journal of Human Resources, 50(2), 317-372.
\item[] Cragg, J. G. and Donald, S. G. (1993). Testing identifiability and specification in instrumental variable models. Econometric Theory, 9(02), 222-240.
\item[] DiTraglia, F. J. (2015). Using invalid instruments on purpose: Focused moment selection and averaging for GMM. Working Paper. University of Pennsylvania.
\item[] Elder, T. E. and Lubotsky, D. H. (2009). Kindergarten entrance age and children’s achievement impacts of state policies, family background, and peers. Journal of Human Resources, 44(3), 641-683.
\item[] Evans, W. N., and Garthwaite, C. (2012). Estimating heterogeneity in the benefits of medical treatment intensity. Review of Economics and Statistics, 94(3), 635-649.
\item[] Evans, W. N. and Lien, D. S. (2005). The benefits of prenatal care: evidence from the PAT bus strike. Journal of Econometrics 125(1), 207-239.
\item[] Guggenberger, P. (2012). On the asymptotic size distortion of tests when instruments locally violate the exogeneity assumption. Econometric Theory 28 (2), 387-421.
\item[] Hahn, J. and Hausman, J. (2005). Estimation with valid and invalid instruments. Annales d'\'{E}conomie et de Statistique, 25-57.
\item[] Hall, A. R. and Inoue, A. (2003). The large sample behavior of the generalized method of moments estimator in misspecified models. Journal of Econometrics 114 (2), 361-394.
\item[] Hall, P. and Horowitz, J. L. (1996). Bootstrap critical values for tests based on generalized-method-of-moments estimators. Econometrica, 64(4), 891-916.
\item[] Heckman, J. J. and Vytlacil, E. (2005). Structural Equations, treatment Effects, and econometric policy evaluation. Econometrica, 73(3), 669-738.
\item[] Imbens, G. W. and Angrist, J. D. (1994). Identification and estimation of local average treatment effects. Econometrica 62(2), 467-475.
\item[] Kitagawa, T. (2015). A Test for Instrument Validity. Econometrica, 83: 2043-2063.
\item[] Koles\'{a}r, M. (2013). Estimation in an instrumental variables model with treatment effect heterogeneity. unpublished manuscript. Cowles Foundation.
\item[] Newey, W. K. and McFadden, D. (1994). Large sample estimation and hypothesis testing. Handbook of Econometrics, 4, 2111-2245.
\item[] Lee, S. (2014). Asymptotic refinements of a misspecification-robust bootstrap for generalized method of moments estimators. Journal of Econometrics, 178(3), 398-413.
\item[] Lochner, L. and Moretti, E. (2004). The effect of education on crime: Evidence from prison inmates, arrests, and self-reports. American Economic Review, 94(1), 155-189.
\item[] Otsu, T. (2011). Moderate deviations of generalized method of moments and empirical likelihood estimators. Journal of Multivariate Analysis 102 (8), 1203-1216.
\item[] Siminski, P. (2013). Employment effects of army service and veterans' compensation: evidence from the Australian Vietnam-era conscription lotteries. Review of Economics and Statistics, 95(1), 87-97.
\item[] Stephens, M. Jr. and Yang, D-Y. (2014). Compulsory education and the benefits of schooling. American Economic Review, 104(6), 1777-92.
\item[] Stock, J. H. and Yogo, M. (2005). Testing for weak instruments in linear IV regression. Identification and inference for econometric models: Essays in honor of Thomas Rothenberg.
\item[] Thornton, R. L. (2008). The demand for, and impact of, learning HIV Status. American Economic Review, 1829-1863.
\end{enumerate}

\end{document}